\documentclass{tlp}
\usepackage[T1]{fontenc}
\usepackage{amsmath}
\usepackage[mathscr]{eucal}
\usepackage{amssymb}
\usepackage{tikz}
\usetikzlibrary{shapes}

\newtheorem{defn}{Definition}
\newtheorem{nota}{Notation}
\newtheorem{ex}{Example}
\newtheorem{property}{Property}
\newtheorem{lem}{Lemma}
\newtheorem{prop}{Proposition}
\newtheorem{cor}{Corollary}

\newcommand{\true}{\ensuremath{\texttt{true}}}
\newcommand{\false}{\ensuremath{\texttt{false}}}
\newcommand{\NOT}{\ensuremath{\operatorname{\textit{not\/}}}}

\newcommand{\ORD}{\ensuremath{\mathrm{Ord}}}
\newcommand{\NSET}{\ensuremath{\mathbb{N}}}
\newcommand{\EQ}{\ensuremath{\doteq}}
\newcommand{\DIF}{\ensuremath{\not\doteq}}

\newcommand{\VOCA}{\ensuremath{{\mathscr V}^\star}}
\newcommand{\VOCAB}{\ensuremath{\mathscr V}}

\newcommand{\INFLVOCAB}{\ensuremath{\mathscr{L}_{\omega_1\omega}(\VOCAB)}}
\newcommand{\FINLVOCAB}{\ensuremath{\mathscr{L}_{\omega\omega}(\VOCAB)}}

\newcommand{\SIM}{\ensuremath{\mathord\sim}}
\DeclareMathOperator{\fv}{fv}
\newcommand{\VAR}[1]{\ensuremath{\fv(#1)}}
\newcommand{\NUM}[1]{\ensuremath{\overline{#1}}}
\DeclareMathOperator{\Prd}{Prd}
\newcommand{\PRD}{\ensuremath{\Prd}}
\newcommand{\PRDV}{\ensuremath{\PRD(\VOCA)}}
\newcommand{\PRDVN}[1]{\ensuremath{\PRD(\VOCA,#1)}}
\newcommand{\IND}{\ensuremath{\mathbb{I}(\VOCA)}}
\newcommand{\WORLDS}{\ensuremath{\mathscr W}}
\newcommand{\FLP}{\ensuremath{\mathcal P}}
\newcommand{\STR}[1]{\ensuremath{\mathfrak #1}}
\newcommand{\MODELS}{\ensuremath{\vDash}}
\newcommand{\WMODELS}{\ensuremath{\MODELS_\WORLDS}}
\newcommand{\FORCES}{\ensuremath{\mathbin{\Vdash}}}
\newcommand{\NFORCES}{\ensuremath{\mathbin{\nVdash}}}
\newcommand{\SPE}[1]{\ensuremath{{\scriptstyle\langle{#1}\rangle_{\FLP}}}}
\newcommand{\extends}[3]{\ensuremath{\mathop{\circledcirc_{#1}^{#2}#3}}}
\newcommand{\selects}[3]{\ensuremath{\mathop{\odot_{#1}^{#2}#3}}}
\DeclareMathOperator{\Unif}{Unif}
\newcommand{\UNIF}[2]{\ensuremath{\Unif(#1,#2)}}

\newcommand{\LB}{[\,}
\newcommand{\RB}{\,]}
\newcommand{\BLB}{\mathopen{\,\bigl[}}
\newcommand{\BRB}{\mathclose{\bigr]\,}}
\DeclareMathOperator{\Clf}{Clf}
\newcommand{\CLF}{\ensuremath{\Clf}}
\newcommand{\PLUS}{\ensuremath{\mathbin{+_\Omega}}}
\newcommand{\PLUSNEG}{\ensuremath{\mathbin{+_{\langle-,-\rangle_{\FLP}}}}}
\newcommand{\PLUSDOTNEG}{\ensuremath{\mathbin{+_{\langle\cdot,-\rangle_{\FLP}}}}}
\newcommand{\SELECT}{\ensuremath{\mathbin{|_\Omega}}}
\newcommand{\SELECTNEG}{\ensuremath{\mathbin{|_{\langle-,-\rangle_{\FLP}}}}}

\title[Contextual hypotheses and logic programming]{Contextual hypotheses and\\ semantics of logic programs}
\author[\'Eric A. Martin]{\'ERIC A. MARTIN\\
            School of Computer Science and Engineering \\
	    The University of New South Wales \\
	    UNSW Sydney NSW 2052, Australia \\
	    \email{emartin@cse.unsw.edu.au}}
\submitted{October 12, 2009}
\revised{May 5, 2010}
\accepted{March 28 2011}

\begin{document}
\maketitle

\begin{abstract}
Logic programming has developed as a rich field, built over a logical substratum whose main constituent is a nonclassical form of negation, sometimes coexisting with classical negation. The field has seen the advent of a number of alternative semantics, with Kripke-Kleene semantics, the well-founded semantics, the stable model semantics, and the answer-set semantics standing out as the most successful. We show that all aforementioned semantics are particular cases of a generic semantics, in a framework where classical negation is the unique form of negation and where the literals in the bodies of the rules can be `marked' to indicate that they can be the targets of hypotheses. A particular semantics then amounts to choosing a particular marking scheme and choosing a particular set of hypotheses. When a literal belongs to the chosen set of hypotheses, all marked occurrences of that literal in the body of a rule are assumed to be true, whereas the occurrences of that literal that have not been marked in the body of the rule are to be derived in order to contribute to the firing of the rule. Hence the notion of hypothetical reasoning that is presented in this framework is not based on making global assumptions, but more subtly on making local, contextual assumptions, taking effect as indicated by the chosen marking scheme on the basis of the chosen set of hypotheses. Our approach offers a unified view on the various semantics proposed in logic programming, classical in that only classical negation is used, and links the semantics of logic programs to mechanisms that endow rule-based systems with the power to harness hypothetical reasoning.
\end{abstract}

\begin{keywords}
Kripke-Kleene semantics, answer-set semantics, stable model semantics, well-founded semantics, classical negation, contextual hypotheses, hypothetical reasoning
\end{keywords}

\section{Motivation}

In this paper, we present a small part of a general framework called parametric logic, some of whose concepts have very practical motivations; the notion of \emph{contextual hypothesis}
and the associated notion of \emph{hypothetical reasoning} are two concepts of this kind.
A contextual hypothesis, according to which a condition can be assumed to be true in some specific contexts rather than globally, is important in web search, as the information sought by users occurs in documents and is found to be relevant thanks to the contextual relationships it bears to the input keywords.
Hypothetical reasoning is relevant to the development of
decision support systems. Effective systems must give users the power to explore and experiment so that they can better understand the domain; they must provide the degree of control that users demand, which the type of hypothetical reasoning to be described here offers. We commence the paper will a simple worked example that introduces the practical aspects of the notions of contextual hypothesis and hypothetical reasoning, and motivates the formal material that follows.

Many decision support systems encode expert knowledge as a set of rules. In practice knowledge bases may have to deal with thousands of rules. Keeping that order of magnitude in mind, imagine a toy example of a knowledge base consisting of rules all of the form \emph{if condition$_1$ and \dots and condition$_n$ then conclusion} abstracted as follows.
\begin{align*}
p_0 &\leftarrow p_0& p_1&\leftarrow p_5& p_3&\leftarrow p_0&p_4&\leftarrow p_1\wedge p_2\wedge p_7\\
p_5&\leftarrow p_4& p_6&\leftarrow p_3\wedge p_4\wedge p_7&p_7&\leftarrow p_6&p_8&\leftarrow p_4\\
p_9&\leftarrow p_5\wedge p_7\wedge p_8&p_{10}&\leftarrow p_3\wedge p_6 
\end{align*}
We could consider a more general set of rules in which some conclusions would be associated with more than one conjunction of conditions, but that would not bring any additional insight. Some of the conditions and conclusions could also carry out negative information, hence be of the form $\neg p$; this example only uses positive information in order to simplify notation, at no conceptual
cost since what will be said of the previous set of rules would be said
mutatis mutandis of a set of rules that also encodes
negative information (the fact that we treat positive and negative information similarly
is one of the hallmarks of our approach, as will be seen across the whole paper).
The above set of rules can be represented by the following diagram, that can be read as a boolean circuit with nothing but \emph{and} gates (with a more general set of rules, we would also have \emph{or} gates, and some conditions would be preceded by a \emph{not} gate).
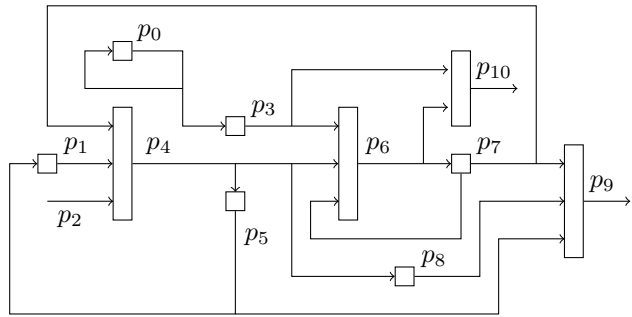
\begin{center}
\begin{tikzpicture}[every node/.style={rectangle,minimum height=0.25cm,minimum width=0.25cm}, every path/.style={->,draw}]
\node[draw] (0) at (1,2.5) {};
\node[draw] (1) at (0,1) {};
\node[draw] (3) at (2.5,1.5) {};
\node[draw, minimum height=1.5cm] (4) at (1,1) {};
\node[draw] (5) at (2.5,0.5) {};
\node[draw, minimum height=1.5cm] (6) at (4,1) {};
\node[draw] (7) at (5.5,1) {};
\node[draw] (8) at (4.75,-0.5) {};
\node[draw, minimum height=1.5cm] (9) at (7,0.5) {};
\node[draw, minimum height=1cm] (10) at (5.5,2) {};

\draw (0) -- node[above, pos=0.35] {$p_0$} (1.8,2.5) -- (1.8,1.5) -- (3);
\draw (1.8,2) -- (0.5,2) -- (0.5,2.5) -- (0);
\draw (1) -- node[above, pos=0.35] {$p_1$} (4);
\draw (0,0.5) -- node[below, pos=0.35] {$p_2$} (0.875,0.5);
\draw (3) -- node[above, pos=0.2] {$p_3$} (3.875,1.5);
\draw (3.25,1.5) -- (3.25,2.25) -- (5.325,2.25);
\draw (4) -- node[above, pos=0.13] {$p_4$} (6);
\draw (3.25,1) -- (3.25,-0.5) -- (8);
\draw (8) -- node[above, pos=0.3] {$p_8$} (5.75,-0.5) -- (5.75,0.5) -- (6.875,0.5) ;
\draw (2.5,1) -- (2.5,0.625);
\draw (5) -- node[right, pos=0.25] {$p_5$} (2.5,-1) -- (-0.5, -1) -- (-0.5,1) -- (1);
\draw (2.5,-1) -- (6,-1)  -- (6,0) -- (6.875,0);
\draw (6) -- node[above, pos=0.22] {$p_6$} (7);
\draw (5,1) -- (5,1.75) -- (5.325,1.75);
\draw (7) -- node[above, pos=0.2] {$p_7$} (6.875,1);
\draw (7) -- (5.5,0) -- (3.5,0) -- (3.5,0.5) -- (3.875,0.5);
\draw (6.5,1) -- (6.5,3.1) -- (0,3.1) -- (0,1.5) -- (0.875,1.5);
\draw (10) -- node[above, pos=0.5] {$p_{10}$} (6.25,2);
\draw (9) -- node[above, pos=0.4] {$p_9$} (7.75,0.5);
\end{tikzpicture}
\end{center}
An important feature of the circuit is that it models a reactive process: it contains a number of loops, which reveal circular arguments. This does not necessarily indicate that the representation of knowledge is flawed. For instance, the deflationary economic model posits that a drop in prices delays consumption, that increases inventories, that forces companies to sell their stock at a lower price. Knowledge representation, when applied to domains where amplifiers and reinforcement mechanisms are at work, usually results in knowledge bases with loops.

In order to establish a correct diagnosis or take the right course of actions, a user will often want to query the decision support system on how a given conclusion can be derived; logically speaking, this is a form of \emph{abductive reasoning}. Getting back to our example, let us query how $p_9$ can be derived. The system could provide a (possibly  minimal) set $X$ of pieces of information that, added to the set of rules, permit $p_9$ to be derived; the members of $X$ would then be assumed to be true \emph{globally}. But the system can do better. It can provide a (possibly minimal) set of occurrences of conditions $X$ in some rules that, supposed to be true at those locations and at those locations only, make the resulting, stronger set of rules able to derive $p_9$; the members of $X$ are then assumed to be true \emph{locally}. Moreover, under this scenario, the system can also indicate which conditions can be \emph{confirmed} (inferred alongside $p_9$). With our running example, this can be done in many different ways. For instance, using check marks to indicate which occurrences of conditions to select and imposing that they be minimal, the system could return
\begin{align*}
p_0 &\leftarrow p_0& p_1&\leftarrow\underset{\checkmark}{p_5}& p_3&\leftarrow p_0&p_4&\leftarrow p_1\wedge\underset{\checkmark}{p_2}\wedge\underset{\checkmark}{p_7}\\
p_5&\leftarrow p_4& p_6&\leftarrow\underset{\checkmark}{p_3}\wedge p_4\wedge\underset{\checkmark}{p_7}&p_7&\leftarrow p_6&p_8&\leftarrow p_4\\
p_9&\leftarrow p_5\wedge p_7\wedge p_8&p_{10}&\leftarrow p_3\wedge p_6 
\end{align*}
and indicate that making $p_2$, $p_3$, $p_5$ and $p_7$ true at the selected locations allows one to infer $p_9$ and confirm $p_5$ and $p_7$, but neither $p_2$ nor $p_3$. Or it could return
\begin{align*}
p_0 &\leftarrow p_0& p_1&\leftarrow p_5& p_3&\leftarrow p_0&p_4&\leftarrow p_1\wedge \underset{\checkmark}{p_2}\wedge p_7\\
p_5&\leftarrow\underset{\checkmark}{p_4}& p_6&\leftarrow\underset{\checkmark}{p_3}\wedge\underset{\checkmark}{p_4}\wedge p_7&p_7&\leftarrow\underset{\checkmark}{p_6}&p_8&\leftarrow p_4\\
p_9&\leftarrow p_5\wedge p_7\wedge p_8&p_{10}&\leftarrow p_3\wedge p_6 
\end{align*}
and indicate that making $p_2$, $p_3$, $p_4$ and $p_6$ true at the selected locations allows one to infer $p_9$ and confirm $p_4$ and $p_6$, but neither $p_2$ nor $p_3$. Since they are not confirmed, $p_2$ and $p_3$ make it possible to derive $p_9$ by playing a `foundational' role, and their marked occurrences indicate where that role is played in the underlying derivation of $p_9$. On the other hand, $p_5$ and $p_7$ (first marked set of rules), or $p_4$ and $p_6$ (second marked set of rules), being confirmed, make it possible to derive $p_9$ thanks to relationships of `interdependence', and their marked occurrences indicate which rules use them as hypotheses in the underlying derivation of $p_9$ before these relationships take effect and the hypotheses become unnecessary as they get confirmed.\footnote{The minimality constraint is essential in this discussion: if the rules were $q_3\leftarrow q_2$ and $q_2\leftarrow q_1$, the aim was to derive $q_3$, and the system returned $q_3\leftarrow\underset{\checkmark}{q_2}$ and $q_2\leftarrow\underset{\checkmark}{q_1}$, then $q_2$ would be confirmed though $q_2$ is not in a relationship of interdependence to itself.} An output of this kind is of great interest to users who, given a confirmed selected condition $\varphi$, can investigate further the feedbacks in which $\varphi$ is involved, which might result in valuable findings or prompt users to amend the knowledge base---they will be all the more prepared to this eventuality the number of rules is larger. Of course, a well designed system will assist in this task.

Other scenarios of interest are possible. Users could first select some occurrences of conditions to indicate the contexts in which those conditions can be assumed to be true, and then list some of those conditions. For instance, users could select, in some rules, some occurrences of the conditions $p_1$, $p_5$ and $p_6$, before listing successively $p_1$, then $p_1$ and $p_5$, then $p_1$ and $p_6$, then $p_1$, $p_5$ and $p_6$, to `activate' first only some, and eventually all, selected occurrences of conditions, and find out what the implications are, what can or cannot be derived as more or fewer assumptions are made in the preselected contexts. Or users could first list conditions, say $p_1$, $p_5$ and $p_6$, and then experiment by selecting various occurrences of those conditions, starting for instance with all of them (so ignoring the context), and then removing some occurrences, hence taking the context into account to find out how that affects the conclusions that can be derived, or the conditions that can be confirmed. Extra constraints can be imposed on which conditions should be confirmed or not, or on the relationships between confirmed conditions, etc.

What does all this have to do with the semantics of logic programs, which is what this paper focuses on? Well, we will see that the notions of hypothetical reasoning and contextual hypothesis can do more than enrich the field of logic programming. They allow one to look at its fundamental semantics from a novel perspective. We will see that these fundamental semantics can be all unified under the umbrella of contextual, hypothetical reasoning. They correspond to particular, highly constrained, ways of selecting occurrences of conditions and of choosing hypotheses. The notion of confirmation plays a pivotal role in one semantics (the well-founded semantics). In the previous example, conditions and conclusions were all positive, making it impossible to derive a contradiction. When conditions and conclusions can be negative, a notion of \emph{nonrefutation} naturally enters the stage to express that contextually hypothesising $p$ does not allow one to infer $\neg p$, or that contextually hypothesising $\neg p$ does not allow one to infer $p$. The notion of nonrefutation plays a crucial role in other semantics (the stable model and the answer-set semantics). Revisiting the fundamental semantics of logic programs under the light of hypothetical reasoning and contextual hypothesis is of conceptual and theoretical interest; in particular, it supports the view that, in contrast to the traditional work in the field, logic programming does not need nonclassical negation, and that positive and negative information can obey a duality principle. As importantly, we can capitalise on the fact that the traditional semantics have been extensively studied and are very well understood, and be confident that expressing them as particular forms of hypothetical reasoning will help understand the latter and come up with valuable constraints and fruitful strategies to exploit it fully.

We have introduced the key idea of choosing a set $X$ of conditions and contextually selecting in some rules some occurrences of conditions, that will be made `active' if they belong to $X$. Formalising this idea precisely, omitting no detail, in a very general setting (in particular because it is first-order rather than propositional) requires a bit of work, but the informal description where check marks are used to capture the notion of contextual hypothesis actually says it all and does not hide any essential technicality. We will use this description again. It does not only support the intuition and illustrate the mathematical developments. It actually suggests a very practical, concrete interface, where users click on some occurrences of conditions, activating or deactivating them as they interact with their decision support system. But users will need to be guided, they should not click arbitrarily, they will need and request some constraints that will help them experiment effectively and beneficially. One can imagine for instance that some occurrences of conditions are dimmed out in real time to indicate that under the present circumstances, they are `unclickable'. This paper will not dwell into these considerations though; the first task is to provide the theoretical foundations to practical problems of this kind.

\section{Background}\label{ba}

Since its inception, the field of logic programming has embraced an increasingly complex diaspora of rules, that is, pairs of formulas referred to as `body' and `head', with the intention that if the body is true then the head is true. The bodies of the rules have eventually been allowed to contain both classical negation and nonclassical negation---classical negation being used to assert falsity, and nonclassical negation, a form of nonprovability.
Classical negation and disjunction have made their ways into the heads of the rules (see~\cite{Minker:2002} for a survey). It has even been advocated to use more than two kinds of negation~\cite{Alferes:1996,Alferes:1998}. Also, a large number of constraints on the rules that make up a logic program have been proposed, based on syntactic constraints or definability properties (\emph{e.g.}, \cite{Jager:1993}) or on proof-theoretic criteria (\emph{e.g.}, \cite{Pedreschi:2002}). All these developments took place as part of the advances in the field of nonmonotonic reasoning~\cite{Minker:1993}. 

Starting with the simplest case of sets of rules whose heads are atomic formulas and whose bodies result from the application of conjunction and disjunction to atomic formulas only, a recurring question has been: what is the intended meaning of a set of rules, that translates into: what are the intended interpretations of a set of rules? Some approaches seek a unique intended interpretation, while other approaches accommodate many.
In a first-order setting, the intended interpretations have been selected from the class of all structures or from the more restricted class of all Herbrand structures, that give every individual a unique name. Alongside the various model-theoretic semantics, proof-theoretic techniques and fixed-point constructions have been developed (see~\cite{Apt:1994} for a survey). As the number of approaches increased, a natural line of research has been to exhibit possible relationships between the various frameworks and seek unifications, with~\cite{Loyer:2003} and~\cite{Hitzler:2005} as examples of work conducted in the last decade. This study belongs to that category of papers, but differs from previous work in many essential ways.
\begin{itemize}
\item
It offers a model of hypothetical reasoning for knowledge-based systems where a hypothesis is not conceived of globally as a new fact, but as a statement meant to be assumed locally and contextually (at some locations in the bodies of some rules), that can be subjected to confirmation (if that statement is eventually  derived), or subjected to not being refuted (if no rule ever produces its negation), or subjected to other constraints, possibly involving the whole set of chosen hypotheses, or some particular subset.
\item
Rather than seeking relationships between various semantics of logic programs, it proposes a natural base semantics, and complements it with a generic notion of transformation of a logic program. So rather than proposing, for a given logic program \FLP, a picture of the form
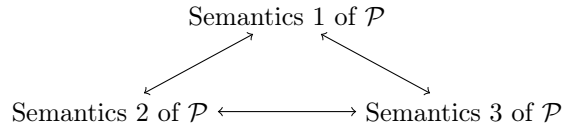
\begin{center}
\begin{tikzpicture}
\matrix [row sep=0.8cm,column sep=-0.5cm] {
& \node (a) {Semantics 1 of \FLP}; \\
\node (b) {Semantics 2 of \FLP}; & & \node (c) {Semantics 3 of \FLP}; \\
};
\draw[<->] (a) -- (b);
\draw[<->] (a) -- (c);
\draw[<->] (b) -- (c);
\end{tikzpicture}
\end{center}
it proposes a picture of the form
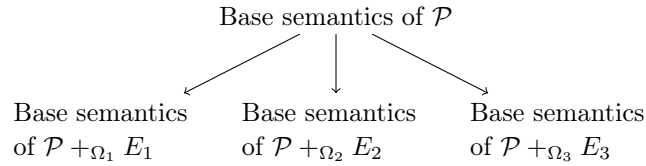
\begin{center}
\begin{tikzpicture}
\matrix [row sep=0.8cm] {
& \node (a) {Base semantics of \FLP}; \\
\node[text width=2.5cm] (b) {Base semantics of $\FLP+_{\Omega_1}E_1$}; & \node[text width=2.5cm] (c) {Base semantics of $\FLP+_{\Omega_2}E_2$}; & \node[text width=2.5cm] (d) {Base semantics of $\FLP+_{\Omega_3}E_3$}; \\
};
\draw[->] (a) -- (b);
\draw[->] (a) -- (c);
\draw[->] (a) -- (d);
\end{tikzpicture}
\end{center}
where $\FLP\PLUS E$ represents the transformation of \FLP\ into a new logic program the bodies of whose rules are possibly weaker than the bodies of the corresponding rules of \FLP, thanks to a construction that uses a set $E$ of literals conceived of as potential hypotheses and a set $\Omega$ of occurrences of literals in the bodies of \FLP's rules conceived of as possible targets of the hypotheses (these notions and others used in the semi-formal presentation of Section~\ref{ba}  will be precisely defined from Section~\ref{lb} onwards). Semantics 1, 2 and 3 of \FLP\ then correspond to particular choices of $E$ and $\Omega$, and intuitively receive the interpretation: \emph{in the bodies of \FLP's rules, make use of the hypotheses in $E$ locally and contextually as indicated by $\Omega$, and apply the base semantics}.
\item
It attains a high degree of unification between the semantics of logic programs considered in this paper, namely, Kripke-Kleene semantics~\cite{Fitting:1985}, the well-founded semantics~\cite{Gelder:1991}, the stable model semantics~\cite{Gelfond:1988}, and the answer-set semantics~\cite{Gelfond:1991}. Kripke-Kleene semantics is closely related to our base semantics, while each of the other three is obtained by instantiating general principles (constraints on $\Omega$ and $E$) that determine families of semantics. Other families would be determined by other principles. Some members of those families, different to the particular members of the particular families considered here, might be worth investigating and have practical use.
\item
It is classical, in the sense that it uses a unique form of negation, interpreted classically, and is based on interpretations that assign one of the classical truth values of true or false to every formula. This is in contrast to many approaches, for instance frameworks based on Belnap's 4-valued logic (see~\cite{Fitting:1999}), or on the extension of the logic of here-and-there, $N_5$, with its 5 truth values (see~\cite{Pearce:2006}).
\item
Using classical negation as unique form of negation, it remains outside the realm of nonmonotonic reasoning. It is monotone in both $\Omega$ and $E$: increasing the targets of hypotheses or the set of potential hypotheses results in stronger programs, that generate more literals.
\item
It is symmetric, in the sense that it treats negated atoms and atoms on a par, and emphasises that the stable model and well-founded semantics, dedicated to interpreting a nonclassical form of negation, give rise to semantics that are fully biased towards negated atoms, while the general principle underlying these semantics is consistent with being totally biased towards nonnegated atoms, or with being committed to achieving a balance between negated and nonnegated atoms.
\item
It applies to general sets of rules, whose heads can be negated atoms and whose bodies can (but do not have to) be arbitrary infinitary first-order formulas.
\item
It does not require that intended interpretations be restricted to the class of Herbrand interpretations.
\end{itemize}
 
\subsection{Two key principles}

Our framework relies on two key principles.

The first principle is that a set of positive rules, that is, rules whose heads are atomic formulas, can be thought of as a set of rules that are both positive and negative, that is, rules whose heads are atomic formulas or negations of atomic formulas, where the negative rules are left \emph{implicit} because they are fully determined by the positive rules thanks to a duality principle. This idea is far from novel; it is nothing more than a variation on the notion of Clark's completion of a logic program~\cite{Clark:1987}.
Clark's completion does not transform a set of positive rules into a set of positive and negative rules, but rather into a set of logical equivalences augmented with unique name axioms. Our formalisation is a streamlined version of Clark's completion. With positive rules only, one can only infer some negative information by failing to generate some positive information---the process known as negation as finite failure that certainly compels us to adopt the view that negation in logic programming is essentially nonclassical. But given both positive and negative rules, one can generate both positive and negative information, and conceive of negation as finite failure as an ingenious proof technique to generate negative information from the positive rules only, as an alternative to generating negative information using both the positive and the negative rules. This paper will demonstrate that this view is perfectly tenable; classical negation is all one needs, and negation as finite failure can be understood as part of a more general inference mechanism that generates nothing but logical consequences. Not surprisingly, this will result in a semantics which, in case the class of intended interpretations is the class of Herbrand interpretations, is fundamentally equivalent to Kripke-Kleene semantics~\cite{Fitting:1985}. We will not make any restriction on the class of intended interpretations, and present our semantics in the most general setting.

The second principle will allow us to stick to our semantics as `the base semantics', while still accounting for the well-founded semantics, the stable model semantics, and the answer-set semantics. This second principle is based on the idea that any of those semantics `force' some assumptions to be made in some parts of some rules, resulting in a new logic program whose base semantics is precisely the desired semantics of the original program. To force some assumptions to be made in some parts of some rules, we use a particular kind of transformation of a logical formula, that we now introduce. Consider two formulas, $\varphi$, of the form
\[
\exists x\bigl(p(x)\wedge q(x)\bigr)\vee\exists x\bigl(p(x)\wedge r(x)\bigr),
\]
and $\psi$, of the form
\[
\exists x\bigl(p(x)\wedge q(x)\bigr)\vee\exists x\bigl((p(x)\vee x\EQ a\vee x\EQ b)\wedge r(x)\bigr)
\]
where $\doteq$ represents identity (denotation by the same closed term of two individuals).
Then we can read $\psi$ as ``$\varphi$, where the second occurrence of $p(x)$ is assumed to be true in case $x$ is either $a$ or $b$.'' Or to put it another way, if in $\varphi$, we hypothesise that $p(a)$ and $p(b)$ are true in the context given by the second 
occurrence of $p(x)$ in $\varphi$, then we get (a logical representation equivalent to) $\psi$. More generally, we will formalise the notion of `transforming a formula into another by making some contextual hypotheses in the former', similar to the way $\varphi$ can be transformed into $\psi$ by making the hypotheses $p(a)$ and $p(b)$ in the context given by the second occurrence of $p(x)$ in $\varphi$.

Having this notion of `contextual hypothesis' and associated formula transformation in hand, our `classical' approach to logic programming will replace the question of ``what should be acknowledged to fail to be derived from a logic program?''---the question at the heart of the well known semantics in the `nonclassical' approaches to logic programming---by the question of ``what contextual hypotheses should be made in the bodies of the rules of a logic program?'' This will allow us to revisit the main semantics that have been proposed and view them as particular members of families of semantics, and more particularly, as those members that are `maximally biased' towards negative information. For an illustration, consider a vocabulary with a constant \NUM{0}, a unary function symbol $s$ and a unary predicate symbol $p$, and the logic program \FLP\ consisting of the following rule.
\[
p(X) \leftarrow p(s(s(X)))
\]
Given a natural number $n$, write \NUM{n} for the term obtained from \NUM{0} by $n$ applications of $s$. Applied to \FLP, the well-founded semantics makes all of $p(\NUM{n})$, $n\in\NSET$, false in its intended model of \FLP, based on the principle that when a logic program presents an infinite descending chain of atoms, all members of that chain should be set to false. It turns out that this is a particular case of a more general principle, that will be formalised in the body of the paper, consistent with a large number of models of \FLP, including in particular
\begin{itemize}
\item
structures in which $p(\NUM{n})$ is false for all $n$'s;
\item
structures in which $p(\NUM{n})$ is true for all $n$'s;
\item
structures in which $p(\NUM{n})$ is false for all even $n$'s, but true for all odd $n$'s;
\item
structures in which $p(\NUM{n})$ is true for all even $n$'s, but false for all odd $n$'s.
\end{itemize}
So this more general principle isolates a number of Herbrand models one of which is maximally biased towards negative information, that happens to be the intended model advocated by the well-founded semantics; but this more general principle can be instantiated to `cousin semantics' of the well-founded semantics, some of which could be of interest. One could be maximally biased towards positive information---a form of dual well-founded semantics---, or one could try and keep a balance between positive and negative information.

\subsection{A mechanistic view on rules}\label{mvr}

The rules that make up a logic program are expressions of the form
\[
\textsl{head}\leftarrow\textsl{body}
\]
that are read in many possible ways. One can view $\leftarrow$ as a link between cause and effect and conceive of \textsl{body} as a statement that if \emph{activated}, allows the rule to \emph{fire} and \textsl{head} to be \emph{generated}; when formally defined, this amounts to a kind of \emph{operational} semantics. Or one can view $\leftarrow$ as a link between antecedent and consequent and conceive of \textsl{body} as a statement that if \emph{true}, allows the rule to be \emph{logically applicable} and \textsl{head} to be established as \emph{true}; when formally defined, this amounts to a \emph{denotational} semantics. A legitimate aim is to propose both an operational and a denotational semantics, and make sure that they match.
In this paper, we propose an operational semantics as it is the shortest path to casting Kripke-Kleene semantics, the well-founded semantics, the stable model semantics, and the answer-set semantics into our framework. We also have a denotational semantics but will make it the subject of another paper.

Let us specify a bit more the syntactic structure of rules and the process by which they fire.
Recall that a formula is in \emph{negation normal form} if negation is applied to atomic formulas only; so formulas in negation normal form are built from \emph{literals} (atomic formulas and their negations) using disjunction, conjunction, existential quantification, and universal quantification. Assume that every rule $\textsl{head}\leftarrow\textsl{body}$ of a logic program is such that \textsl{head} is a literal and \textsl{body} is a formula in negation normal form. Firing rules causes literals---the heads of the rules that fire---to be \emph{generated}. Literals can be combined into formulas in negation normal form some of which can, thanks to the generated literals, be \emph{inferred}. We impose that inferring formulas in negation normal form be a \emph{constructive} process; so $p\vee\neg p$ can be inferred provided that $p$ or $\neg p$ has been generated, and $\exists x\,p(x)$ can be inferred provided that $p(t)$ has been generated for at least one  closed term $t$.

Having literals as heads of the rules of a logic program is natural in relation to the answer-set semantics. We will see that it is also natural in relation to Kripke-Kleene, the well-founded and the stable model semantics, thanks to the notions of \emph{duality} of a formula and of \emph{symmetry} of a logic program, that we now introduce. Given a formula $\varphi$ in negation normal form, define the dual of $\varphi$ as the formula $\SIM\varphi$ obtained from $\varphi$ by changing disjunction into conjunction, conjunction into disjunction, existential quantification into universal quantification, universal quantification into existential quantification, by negating nonnegated atomic formulas and deleting all negation signs (before atomic formulas). For instance, if $\varphi$ is
\[
(p(X)\vee \neg q(X))\wedge(\neg p(X)\vee r(X))
\]
then the dual $\SIM\varphi$ of $\varphi$ is
\[
(\neg p(X)\wedge q(X))\vee(p(X)\wedge\neg r(X)).
\]
Now say that a logic program \FLP\ is symmetric if the bodies of all rules are formulas in negation normal form and if for all $n\in\NSET$ and $n$-ary predicate symbols $\wp$, \FLP\ contains exactly two rules of respective form $\wp(v_1,\dotsc,v_n)\leftarrow\varphi_\wp^+$ and $\neg\wp(v_1,\dotsc,v_n)\leftarrow\varphi_\wp^-$ that are dual of each other in the sense that $\varphi_\wp^-$ and $\varphi_\wp^+$ are dual of each other. With these notions in hand, we will be able to view Kripke-Kleene semantics, the well-founded semantics and the stable model semantics as applied to symmetric logic programs. The working hypothesis is that all three semantics \emph{do} deal with symmetric logic programs even though traditionally, many rules can have a head built from a given predicate symbol  and only the positive rules are explicitly given, with the negative rules being implicit; this is legitimate as first, negation normal form is not restrictive, second, a straightforward syntactic transformation allows one to merge all rules whose heads are built from a given predicate symbol, and third, every negative rule is perfectly determined by its dual positive rule.

It seems natural to allow rules to fire finitely often only, as this immediately suggests obvious implementations. But we can think theoretically and assume that rules are allowed to fire transfinitely many times---and all fixed point semantics happily go for it~\cite{Emden:1976,Denecker:2001}. So after all rules have fired any finite number of times, they could fire for the $\omega$\nobreakdash-th time, and then for the $(\omega+1)$\nobreakdash-st  time, and then for the $(\omega+2)$\nobreakdash-nd time\dots\ and then for the $(\omega\times 2)$\nobreakdash-nd time, etc.
For instance, if every literal of the form $p(\NUM{n})$, $n\in\NSET$, has been generated at  stage $5n$, and if all individuals in the domains of all possible interpretations are denoted by a term of the form \NUM{n}, then $\forall x\,p(x)$ can be inferred at stage $\omega$, a point from which any rule whose body is $\forall x\,p(x)$ can fire.
Formalising the process by which rules fire transfinitely often determines the set of literals $\mathop{\mathopen{[}}\FLP\RB$ generated by a set \FLP\ of (positive and negative) rules. It is an operational semantics, previously referred to as the base semantics of \FLP. No other semantics will be proposed: what is presented as an alternative semantics of \FLP\ will be viewed as the base semantics of a program obtained from \FLP\ in a particular way, that captures the essence of the alternative semantics and is an instance of a generic class of transformations.

\subsection{Making contextual hypotheses}\label{mch}

Let us describe a bit more precisely the relationships between the base semantics and the well-founded semantics, the stable model semantics and the answer-set semantics. Consider a set $E$ of literals. Also consider a function $\Omega$, defined on the set of bodies of the rules in \FLP, that returns, for the body $\varphi$ of each rule in \FLP, a selected set of occurrences of literals in $\varphi$. We can represent this function graphically using check marks, writing for instance
\[
\bigl(p(X)\vee \underset{\checkmark}{q(X)}\bigr)\wedge\bigl(\underset{\checkmark}{p(X)}\vee\ r(X)\bigr)
\]
to indicate that the selected occurrences of literals in the formula $\varphi$ defined as
\[
\bigl(p(X)\vee q(X)\bigr)\wedge\bigl(p(X)\vee r(X)\bigr)
\]
are the unique occurrence of $q(X)$ and the second occurrence of $p(X)$.
Now with $E$ and $\Omega$ in hand, we define from \FLP\ a new set of rules, denoted $\FLP\PLUS E$, that formalises the intuitive request: ``in the bodies of the rules of \FLP, use $E$ as a set of hypotheses in the contexts indicated by $\Omega$.'' For instance, if \FLP\ contains the rule $R$ defined as
\[
p(X)\leftarrow\bigl(p(X)\vee q(X)\bigr)\wedge\bigl(p(X)\vee r(X)\bigr),
\]
if $E$ is defined as $\{p(\NUM{2n})\mid n\in\NSET\}\cup\{b(\NUM{n})\mid n\in\NSET\}$, and if $\Omega$ selects the unique occurrence of $q(X)$ and the second occurrence of $p(X)$ in the body of $R$ then $\FLP\PLUS E$ will contain a rule that is logically equivalent to
\[
p(X)\leftarrow\bigl(p(X)\vee q(X)\bigr)\wedge\bigl(p(X)\vee\bigvee_{n\in\NSET}X\doteq\NUM{2n}\vee r(X)\bigr)
\]
where $\doteq$ denotes syntactic identity. We will see that in case \FLP\ is symmetric, we can choose $\Omega$ and $E$ in such a way that $\LB\FLP\PLUS E\RB$ captures the well-founded  semantics applied to the positive rules of \FLP; moreover, this choice of $\Omega$ and $E$ is a particular case of choices made according to a simple principle, that happens to be maximally biased towards negative information. Still in case \FLP\ is symmetric, we can also choose $\Omega$ and $E$ in ways such that $\LB\FLP\PLUS E\RB$ are the stable models of the positive rules of \FLP; similarly, these choices of $\Omega$ and $E$ are particular cases of choices made according to a simple principle, that happen to be maximally biased towards negative information. Importantly, these correspondences are between a framework where negation is classical and frameworks where negation is meant not to be classical. The answer-set semantics seems to offer a greater challenge as its syntax accommodates two kinds of negation: $\neg$, meant to be classical, and $\NOT$, meant to be nonclassical. But the correspondence turns out to be easy to establish if one conceives of $\NOT$ as a syntactic variant to $\Omega$. More precisely, conceive of $\NOT\mathit{literal}$ as a request to select $\SIM\mathit{literal}$. Given a set of rules \FLP\ in the bodies of which both $\neg$ and $\NOT$ might occur, with $\NOT$ applied only to atoms and classical negations of atoms, consider the set of rules $\FLP'$ obtained from \FLP\ by replacing all occurrences of $\NOT\mathit{literal}$ with $\SIM\mathit{literal}$ (so only classical negation occurs in $\FLP'$). Then set $\Omega$ to select precisely the occurrences of literals in the bodies of the rules of $\FLP'$ that have replaced an occurrence of an expression of the form $\NOT\mathit{literal}$ in the bodies of the corresponding rules of \FLP.
For instance, if \FLP\ contains the rule
\[
p(X)\leftarrow\bigl(\NOT p(X)\vee q(X)\bigr)\wedge\bigl(\neg p(X)\vee\NOT\neg r(X)\bigr),
\]
then $\FLP'$ will contain the rule
\[
p(X)\leftarrow\bigl(\neg p(X)\vee q(X)\bigr)\wedge\bigl(\neg p(X)\vee r(X)\bigr)
\]
that $\Omega$ will mark as
\[
p(X)\leftarrow\bigl(\underset{\checkmark}{\neg p(X)}\vee q(X)\bigr)\wedge\bigl(\neg p(X)\vee\underset{\checkmark}{r(X)}\bigr).
\]
It is then easy to choose $E$ in ways such that $\LB\FLP'\PLUS E\RB$ are the answer-sets for \FLP\ (one answer-set for each choice of $E$).

\section{Logical background}\label{lb}

\NSET\ denotes the set of natural numbers and \ORD\ the class of ordinals.

\subsection{Syntax}\label{syn}

\begin{defn}\label{defn_1}
A \emph{vocabulary} is a nonempty set of (possibly nullary) function symbols and (possibly nullary) predicate symbols none of which is the distinguished binary predicate symbol \EQ\ (identity), such that if \VOCAB\ contains at least one nonnullary predicate or function symbol then \VOCAB\ contains at least one nullary function symbol.
\end{defn}

As usual, a constant refers to a nullary function symbol.\footnote{Vocabularies that would contain at least one nonnullary predicate or function symbol but no constant would be degenerate in this setting, and are better ruled out, though they would be perfectly legitimate in the usual treatment of first-order logic. But see the discussion at the beginning of Section~\ref{flp} on how full generality is obtained despite the restrictions imposed on vocabularies.}
We will discuss later the distinction between \EQ\ and equality (=), which note can be one of the predicate symbols in a vocabulary. Accepting nullary predicate symbols in vocabularies will allow us to formalise all notions in a setting that can be either purely propositional, or purely first-order, or hybrid.

\begin{nota}\label{nota_2}
When a vocabulary contains the constant \NUM{0} and the unary 
function symbol $s$, we use \NUM{n} to refer to the term obtained from \NUM{0} by $n$ 
applications of $s$.
\end{nota}

\begin{nota}\label{nota_3}
We denote by \VOCAB\ a vocabulary.
\end{nota}

\begin{nota}\label{nota_4}
We fix a countably infinite set of (first-order) variables together with a repetition-free enumeration $(v_i)_{i\in\NSET}$ of this set.
\end{nota}

By \emph{term} we mean term over \VOCAB, built from the function symbols in \VOCAB\ and the members of $(v_i)_{i\in\NSET}$. We say that a term is \emph{closed} if it contains no variable.

\begin{defn}\label{defn_5}
The set \INFLVOCAB\ of (\emph{infinitary}) \emph{formulas} (\emph{over \VOCAB}) is inductively defined as the smallest set that satisfies the following conditions.
\begin{itemize}
\item
All \emph{literals}---\emph{atoms} and \emph{negated atoms}---(\emph{over \VOCAB}), namely, all expressions of the form $\wp(t_1,\dotsc,t_n)$ or $\neg\wp(t_1,\dotsc,t_n)$ where
$n\in\NSET$, $\wp$ is an $n$-ary predicate symbol in $\VOCAB$, and $t_1$, \dots, $t_n$ are terms over \VOCAB, belong to \INFLVOCAB.
\item
If \VOCAB\ contains at least one constant then all \emph{identities} and \emph{distinctions} (\emph{over \VOCAB}), namely, all expressions of the form $t\EQ t'$ or $\neg t\EQ t'$, the latter being usually written $t\DIF t'$, where $t$ and $t'$ are terms over \VOCAB, belong to \INFLVOCAB.
\item
All expressions of the form $\bigvee X$ or $\bigwedge X$ with $X$ a countable set of formulas over \VOCAB, belong to \INFLVOCAB.
\item
All expressions of the form $\exists x\,\varphi$ or $\forall x\,\varphi$ where $x$ is a variable and $\varphi$ is a formula over \VOCAB\ that has $x$ as a free variable, belong to \INFLVOCAB.
\end{itemize}
\end{defn}

A few observations about the definition of \INFLVOCAB\ are in order.
\begin{itemize}
\item
First, negation is assumed to be applicable to atoms only, which amounts to imposing a negation normal form, at no loss of generality. This is technically convenient, and often used in logic programming.
\item
Second, the application of quantifiers is restricted to formulas that have the quantified variables as free variables, and identities and distinctions are ruled out in case \VOCAB\ contains no constant, again at no loss of generality. This is to embed the propositional framework neatly in a first-order setting: if \VOCAB\ consists of (nullary) predicate symbols only then \INFLVOCAB\ is the infinitary propositional language built on \VOCAB.
\item
Third, if we wanted to sometimes restrict some concepts to finite formulas, then we would still be happy with disjunction and conjunction defined as unary operators on finite sets of formulas. This treatment of disjunction and conjunction, which contrasts with the traditional view of binary operators on pairs of formulas, does more than let \INFLVOCAB\ naturally extend the set of finite first-order formulas over \VOCAB. It also simplifies the formal developments. In particular, there is no need to introduce two extra symbols \true\ and \false, as is usually done in logic programming, since $\bigwedge\varnothing$ is valid and can play the role of \true, and $\bigvee\varnothing$ is invalid and can play the role of \false.
\item
Fourth, we distinguish between identity and equality. Identity is treated as a logical symbol, and its interpretation built into the logic, whereas equality is treated as a nonlogical symbol (a possible member of \VOCAB), whose intended interpretation needs to be explicitly provided. This will be discussed at greater length in Section~\ref{flp}.
\end{itemize}
Let us emphasise that our framework does not need the power of infinitary languages. Readers interested only in finite logic programs can ignore the qualifier ``infinitary'' and replace \INFLVOCAB\ by \FINLVOCAB. But \INFLVOCAB\ offers an elegant way to work with logic programs consisting of infinitely many rules built from a vocabulary with a finite number of predicate symbols, possibly obtained by grounding a finite logic program expressed in a first-order language whose set of closed terms is infinite, without making any of the formal notions and proofs more complicated than their finitary counterparts.

Let a formula $\varphi$ be given. We let $\VAR{\varphi}$ denote the set of free variables of $\varphi$. If $\VAR{\varphi}=\varnothing$ then $\varphi$ is said to be \emph{closed}.
Let $e$ be a formula or a term.
Given $n\in\NSET$, distinct variables $x_1$, \dots, $x_n$ and closed terms $t_1$, \dots, $t_n$, we write $e[t_1/x_1,\dotsc,t_n/x_n]$ for the result of substituting simultaneously in $e$ all free occurrences of $x_1$, \dots, $x_n$ by $t_1$, \dots, $t_n$, respectively.
Let $e$ and $e'$ be two formulas or terms. We say that $e'$ is a closed instance of $e$ iff there exists $n\in\NSET$, distinct variables $x_1$, \dots, $x_n$ and closed terms $t_1$, \dots, $t_n$ such that $x_1$, \dots, $x_n$ are the variables that occur free in $e$ and $e'$ is $e[t_1/x_1,\dots,t_n/x_n]$; if $e'$ is known to be closed then we say ``instance of $e$'' rather than ``closed instance of $e$.''
Given $n\in\NSET$ and terms $t_1$, $t'_1$, \dots, $t_n$, $t'_n$, we say that $(t'_1,\dotsc,t'_n)$ is a closed instance of $(t_1,\dots,t_n)$ iff for all members $i$ of $\{1,\dotsc,n\}$, $t'_i$ is a closed instance of $t_i$; when $t'_1$, \dots, $t'_n$ are known to be closed then we say ``instance of $(t_1,\dots,t_n)$'' rather than ``closed instance of $(t_1,\dots,t_n)$.''

Though negation can be applied only to atoms and identities, we need to be able to semantically negate a formula in a syntactically friendly manner which is achieved in the following usual way: given a formula $\varphi$, $\SIM\varphi$ denotes
\begin{itemize}
\item
$\neg\varphi$ if $\varphi$ is an atom;
\item
$\psi$ if $\varphi$ is of the form $\neg\psi$;
\item
$t\DIF t'$ if $\varphi$ is of the form $t\EQ t'$;
\item
$t\EQ t'$ if $\varphi$ is of the form $t\DIF t'$;
\item
$\bigwedge\{\SIM\psi \mid \psi\in X\}$ if $\varphi$ is of the form $\bigvee X$;
\item
$\bigvee\{\SIM\psi \mid \psi\in X\}$ if $\varphi$ is of the form $\bigwedge X$;
\item
$\forall x\,\SIM\psi$ if $\varphi$ is of the form $\exists x\,\psi$;
\item
$\exists x\,\SIM\psi$ if $\varphi$ is of the form $\forall x\,\psi$.
\end{itemize}

Given a set $X$ of formulas, we let $\SIM X$ denote $\{\SIM\varphi\mid\varphi\in X\}$.
A set $X$ of literals is said to be \emph{consistent} just in case there is no closed atom $\varphi$ that is an instance of both a member of $X$ and a member of $\SIM X$. A set of literals is said to be \emph{inconsistent} iff it is not consistent.
A set $X$ of literals is said to be \emph{saturated} just in case every closed atom is an instance of a member of at least one of the sets $X$ and $\SIM X$. A set of literals is said to be \emph{complete} just in case it is saturated and consistent.

Given $n\in\NSET$ and formulas $\varphi_1$, \dots, $\varphi_n$,
we use $\varphi_1\vee\dotsb\vee\varphi_n$ and $\varphi_1\wedge\dotsb\wedge\varphi_n$ as abbreviations for
$\bigvee\{\varphi_i \mid 1\leq i\leq n\}$ and $\bigwedge\{\varphi_i \mid 1\leq i\leq n\}$, respectively.
Also, given two formulas $\varphi_1$ and $\varphi_2$,
$\varphi_1\rightarrow\varphi_2$ is an abbreviation for $\SIM\varphi_1\vee\varphi_2$
and $\varphi_1\leftrightarrow\varphi_2$ is an abbreviation for $(\varphi_1\rightarrow\varphi_2)\wedge(\varphi_2\rightarrow\varphi_1)$. Note that $\rightarrow$ is a logical symbol whereas $\leftarrow$ is not: $\leftarrow$ has been used before and will be used later to represent rules in the traditional way, separating the head of a rule from its body. In the operational semantics that is the subject of this paper, $\leftarrow$ is not meant to be interpreted as logical implication.

The subformulas of a formula $\varphi$ of the form $\bigvee X$ or $\bigwedge X$ are $\varphi$ and the subformulas of the members of $X$.
The subformulas of a formula $\varphi$ of the form $\exists x\,\psi$ or $\forall x\,\psi$ are $\varphi$ and the subformulas of $\psi$.
The subformulas of a formula of the form $\neg\psi$ are $\neg\psi$ and $\psi$.
An atom or identity is its unique subformula.

Let a formula $\varphi$ be given. Let $T$ be the parse tree of $\varphi$ 
where the nodes are labeled with one of $\bigvee$, $\bigwedge$, $\exists x$ for some variable $x$, or $\forall x$ for some variable $x$, so 
that the leaves of $T$ are all (intuitive) occurrences of literals, 
identities and distinctions in $\varphi$. Then a (formal) occurrence of a literal in $\varphi$ 
can be defined as the set of all formulas that appear on the branch of $T$ 
whose leaf is that (intuitive) occurrence of literal.

\begin{defn}\label{defn_6}
Given a formula $\varphi$ and a literal $\psi$, an
\emph{occurrence of\/ $\psi$ in\/ $\varphi$} is defined as a set $O$ of formulas that contains both $\varphi$ and $\psi$ and such that:
\begin{itemize}
\item
all members of $O$ are subformulas of $\varphi$;
\item
$\psi$ is a subformula of all members of $O$;
\item
for all members $\psi_1$ and $\psi_2$ of $O$, either $\psi_1$ is a subformula of $\psi_2$ or $\psi_2$ is a subformula of $\psi_1$; 
\item
for all members of $O$ of the form $\bigvee X$ or $\bigwedge X$, $O$ contains a member of $X$;
\item
for all members of $O$ of the form $\exists x\,\xi$ or $\forall x\,\xi$, $O$ contains $\xi$.
\end{itemize}
\end{defn}

\begin{ex}\label{ex_7}
Suppose that \VOCAB\ contains 3 nullary predicate symbols $p$, $q$ and $r$. Let $\varphi$ 
denote $\bigwedge\bigl\{\neg p,\,\bigvee\{q,\,r,\,\neg p\}\bigr\}$. The occurrences of literals in $\varphi$ are:
\begin{itemize}
\item
$\{\varphi,\,\neg p\}$---an occurrence of $\neg p$ in $\varphi$;
\item
$\bigl\{\varphi,\,\bigvee\{q,\,r,\,\neg p\},\,q\bigr\}$---an occurrence of 
$q$ in $\varphi$;
\item
$\bigl\{\varphi,\,\bigvee\{q,\,r,\,\neg p\},\,r\bigr\}$---an occurrence of 
$r$ in $\varphi$;
\item
$\bigl\{\varphi,\,\bigvee\{q,\,r,\,\neg p\},\,\neg p\bigr\}$---an 
occurrence of $\neg p$ in $\varphi$.
\end{itemize}
\end{ex}

\subsection{Semantics}

\begin{defn}\label{defn_8}
Let a set $S$ of literals be given.

For all formulas $\varphi$, we inductively define the notion \emph{$S$ forces\/ $\varphi$}, denoted $S\FORCES\varphi$, as follows.
If $S$ is inconsistent then $S$ forces all formulas.
Assume that $S$ is consistent.
\begin{itemize}
\item
For all formulas $\varphi$, $S\FORCES\varphi$ iff $S$ forces all closed instances of $\varphi$.
\item
For all closed terms $t_1$ and $t_2$, $S\FORCES t_1\EQ t_2$ in case $t_1$ and $t_2$ are 
identical, and $S\FORCES t_1\DIF t_2$ in case $t_1$ and $t_2$ are distinct.
\item
For all closed literals $\varphi$, $S\FORCES\varphi$ iff $\varphi$ is an instance of a member of $S$.
\item
For all countable sets $X$ of closed formulas, $S\FORCES\bigvee X$ iff $S$ forces some member of $X$, and $S\FORCES\bigwedge X$ iff $S$ forces all members of $X$.
\item
For all formulas $\varphi$ and variables $x$ with $\VAR{\varphi}=\{x\}$,
$S\FORCES\exists x\,\varphi$ iff $S\FORCES\varphi[t/x]$ for some closed term $t$,
and $S\FORCES\forall x\,\varphi$ iff $S\FORCES\varphi[t/x]$ for all closed terms $t$.
\end{itemize}

Given a set $T$ of formulas, we say that \emph{$S$ forces\/ $T$}, denoted $S\FORCES T$, just in case $S$ forces all members of $T$.
\end{defn}

\begin{defn}\label{defn_9}
A \emph{standard structure} (\emph{over\/ \VOCAB}) is a set of closed atoms.
\end{defn}

Note the following particular cases.
\begin{itemize}
\item
If \VOCAB\ contains no nullary predicate symbol then a standard 
structure over \VOCAB\ is basically a Herbrand interpretation.
\item
If \VOCAB\ contains (nullary) predicate symbols only then a standard structure is basically a 
propositional interpretation.
\end{itemize}

The following is the usual notion of truth of a formula in a structure, concisely expressed in terms of the notion introduced in Definition~\ref{defn_8}, which of course is meant to serve other purposes.

\begin{defn}\label{defn_10}
Let a standard structure \STR{M} be given. Let $X$ be the complete set of closed literals such that for all closed atoms $\varphi$, $\varphi\in X$ iff $\varphi\in\STR{M}$.
For all formulas $\varphi$, we say that \emph{$\varphi$ is true in\/ \STR{M}}, or that \emph{\STR{M} is a model of\/ $\varphi$}, iff $X\FORCES\varphi$.
\end{defn}

\begin{nota}\label{nota_11}
Let a standard structure \STR{M} be given. Given a formula $\varphi$, we write $\STR{M}\MODELS\varphi$ if \STR{M} is a model of $\varphi$, and $\STR{M}\not\MODELS\varphi$ otherwise.  Given a set $T$ of formulas, we write $\STR{M}\MODELS T$ if \STR{M} is a model of all members of $T$, and $\STR{M}\not\MODELS T$ otherwise.
\end{nota}

\begin{nota}\label{nota_12}
We denote by \WORLDS\ the set of all standard structures (over \VOCAB).
\end{nota}

Given a set $T$ of formulas and a formula $\varphi$, we write $T\WMODELS\varphi$ if every standard model of \ $T$ is a model of $\varphi$;
if $T\WMODELS\varphi$ then we say that \emph{$T$ logically implies\/ $\varphi$ 
in\/ \WORLDS} or that \emph{$\varphi$ is a logical consequence of\/ $T$ in\/ \WORLDS}.
The same notation and terminology also applies to sets of formulas.
Two formulas $\varphi$ and $\psi$ are said to be \emph{logically equivalent in\/ \WORLDS} iff they have the same models in \WORLDS.

\section{Formal logic programs and their denotational semantics}

\subsection{Formal logic programs}\label{flp}

The concepts introduced in the previous section might suggest that we are considering a notion of logical consequence, namely $\MODELS_\WORLDS$, that, because of its focus on standard structures, is stronger than the classical notion of logical consequence. To make sure that this is not necessarily the case and achieve full generality, we distinguish between two kinds of vocabularies, namely, a vocabulary meant to \emph{describe} a structure and a vocabulary meant to \emph{talk about} a structure. The vocabulary \VOCAB\ introduced in Notation~\ref{nota_3} is of the first kind; it is meant to express what a structure is `made of', but it might not be the vocabulary used to \emph{talk about} a structure, to express properties of a structure. We assume that the vocabulary used to talk about a structure is no more expressive, and is possibly less expressive, than the vocabulary used to describe a structure.

\begin{nota}\label{nota_13}
We denote by \VOCA\ a (possibly finite) countable subset of \VOCAB.
\end{nota}
 
\VOCA\ is the vocabulary to be used when we talk about a structure by writing down theories, axioms, theorems: all must consist of formulas over \VOCA. Suppose that infinitely many closed terms are not terms over \VOCA, either because \VOCAB\ contains infinitely many constants not in \VOCA, or because \VOCAB\ contains at least one function symbol of arity one or more that is not in \VOCA. Then for all sets $T$ of formulas over \VOCA\ and for all formulas $\varphi$ over \VOCA, $T\WMODELS\varphi$ iff $T\MODELS\varphi$. In other words, if countably many closed terms are `unspeakable of' then $\WMODELS$, with sets of formulas that can be `spoken out' on the left hand side and with formulas that can be `spoken out' on the right hand side, is equivalent to the classical notion of logical consequence~\cite{Martin:2006}. This means that by choosing \VOCAB\ to be countable and by setting \VOCA\ to \VOCAB, one opts for a semantics based on Herbrand structures, but by setting \VOCA\ to a strict subset of \VOCAB\ that makes countably many closed terms `unspeakable of', then one opts for a semantics equivalent to the classical notion of logical consequence defined on the basis of all structures.

The availability of both \VOCAB\ and \VOCA\ therefore provides a uniform and simple way to express that a result holds for the classical notion of logical consequence as well as for the more restricted notion of logical consequence that rules out nonstandard structures---by not imposing any condition on \VOCA\ in the statement of that result---, or to force a result to hold for one notion of logical consequence only---by imposing the right condition on the relationship between \VOCAB\ and \VOCA. Restricting in different ways a given vocabulary offers some advantages over the more traditional approach of expanding in different ways a given vocabulary, as done in the Henkin proof of the completeness of first-order logic or in so-called Herbrand semantics of first-order logic (see for instance~\cite{Kaminski:2002}).

Most of the work done in logic programming is developed on the basis of the class of Herbrand structures. But there are exceptions, for instance, the semantics of definite logic programs and queries can be based on either Herbrand structures or all structures: given a definite logic program $T$ and a definite query $Q$, Prolog returns a computed answer substitution $\theta$ iff the universal closures of $Q\theta$ are true in all Herbrand models of $T$, or equivalently, are true in the minimal Herbrand model of $T$, or equivalently, are true in all models of $T$~\cite{Lloyd:1987}. So it is sometimes desirable not to be tied to a semantics based on Herbrand structures.
Moreover, we will see that such a restriction is not conceptually necessary in the sense that all notions defined in this paper will not require any prior assumption on the relationship between \VOCA\ and \VOCAB; but we will sometimes have to suppose that \VOCA\ is equal to \VOCAB\ in the statements of some results. 
So we are going to define a notion of logic program as a set of rules built from \VOCA, not from \VOCAB.
 
\begin{nota}\label{nota_14}
We denote by \PRDV\ the set of predicate symbols in \VOCA.
For all $n\in\NSET$, we denote by \PRDVN{n} the set of members of \PRDV\ of arity $n$.
\end{nota}

We want to consider sets of rules whose heads are literals and whose bodies are arbitrary.
Since formulas can be infinitary and can contain occurrences of $\doteq$, and since $\bigvee\varnothing$ is a formula that can be used as the body of a rule such as $q\leftarrow\bigvee\varnothing$ to express that $q$ is neither a fact nor the head of a rule that can be activated, it is enough to provide, for every $n\in\NSET$ and $\wp\in\PRDVN{n}$, two rules: one whose head is $\wp(v_1,\dotsc,v_n)$ (which is nothing but $\wp$ if $n=0$), and one whose head is $\neg\wp(v_1,\dotsc,v_n)$ (which is nothing but $\neg\wp$ if $n=0$). For instance,
$\{p_1(\NUM{2n})\leftarrow p_2(\NUM{2n+1})\mid n\in\NSET\}$
can be represented as
$p_1(v_1)\leftarrow\textstyle\bigvee\bigl\{p_2(s(v_1))\wedge v_1\doteq \NUM{2n}\bigm| n\in\NSET\bigr\}$.

For the purpose of providing, for every $\wp\in\PRDV$, the positive and negative rules associated with $\wp$, and for the purpose of making some definitions more compact, we introduce the following notation (in which $+$ and $-$ could be replaced by 1 and 0, but using $+$ and $-$ will be easier to read).

\begin{nota}\label{nota_15}
We let \IND\ denote $\PRDV\times\{+,-\}$.

Given $n\in\NSET$, $\wp\in\PRDVN{n}$ and terms $t_1$, \dots, $t_n$, we also write $\wp(t_1,\dotsc,t_n)$ as $\wp^+(t_1,\dotsc,t_n)$ and $\neg\wp(t_1,\dotsc,t_n)$ as $\wp^-(t_1,\dotsc,t_n)$.
\end{nota}

\begin{defn}\label{defn_16}
We define a \emph{formal logic program} (\emph{over\/ \VOCA}) as an \IND-family of formulas over \VOCA, say $(\varphi_\wp^\epsilon)_{(\wp,\epsilon)\in\IND}$, such that for all $n\in\NSET$, $\wp\in\PRDVN{n}$ and $\epsilon\in\{+,-\}$, $\VAR{\varphi_\wp^\epsilon}\subseteq\{v_1,\dotsc,v_n\}$.
\end{defn}

Since there is no restriction on the use of quantifiers in the body of a rule, the condition on variables is at no loss of generality and is imposed so as to simplify subsequent notation, and is also often used in the literature; it just states that a variable that occurs free in the body of a rule occurs in the head of that rule.
Note that if the set of predicate symbols in \VOCA\ is finite then finite sets of finite rules over \VOCA\ are naturally translated into finite formal logic programs.

\begin{ex}\label{ex_17}
Suppose that \VOCA\ consists of a constant \NUM{0}, a unary function symbol $s$, 5 nullary predicate symbols $q_1$, \dots, $q_5$, and 4 unary predicate symbols $p_1$, \dots, $p_4$. The following formulas provide an example of a formal logic program.
\begin{align*}
\varphi^+_{p_1}&\equiv v_1\doteq\NUM{0}\vee\exists v_0\bigl(v_1\doteq s(s(v_0))\wedge p_1(v_0)\bigr) \\
\varphi^-_{p_1}&\equiv  v_1\not\doteq\NUM{0}\wedge\forall v_0\bigl(v_1\not\doteq s(s(v_0))\vee\neg p_1(v_0)\bigr) \\[2mm]
\varphi^+_{p_2}&\equiv v_1\doteq\NUM{0}\vee\exists v_0\bigl(v_1\doteq s(s(v_0))\wedge p_2(v_0)\bigr) \\
\varphi^-_{p_2}&\equiv v_1\doteq s(\NUM{0})\vee\exists v_0\bigl(v_1\doteq s(s(v_0))\wedge\neg p_2(v_0)\bigr) \\[2mm]
\varphi^+_{p_3}&\equiv v_1\doteq\NUM{0}\vee\exists v_0\bigl(v_1\doteq s(v_0)\wedge\neg p_3(v_0)\bigr) \\
\varphi^-_{p_3}&\equiv \exists v_1\bigl(v_1\doteq s(v_0)\wedge p_3(v_0)\bigr) \\[2mm]
\varphi^+_{p_4}&\equiv p_4(s(s(v_1)))& \varphi^+_{q_1}&\equiv \textstyle\bigwedge\varnothing& \varphi^+_{q_2}&\equiv q_3 \\
\varphi^-_{p_4}&\equiv \neg p_4(s(s(v_1)))& \varphi^-_{q_1}&\equiv \textstyle\bigvee\varnothing& \varphi^-_{q_2}&\equiv \neg q_3 \\[2mm]
\varphi^+_{q_3}&\equiv q_2& \varphi^+_{q_4}&\equiv q_4& \varphi^+_{q_5}&\equiv \neg q_5 \\
\varphi^-_{q_3}&\equiv \neg q_2& \varphi^-_{q_4}&\equiv \textstyle\bigvee\varnothing& \varphi^-_{q_5}&\equiv \textstyle\bigvee\varnothing
\end{align*}
\end{ex}

As \VOCA\ contains both nullary and nonnullary predicate symbols, Example~\ref{ex_17} describes a `hybrid' formal logic program, but of a simple kind as it consists of a purely first-order part and a purely propositional part. Let us take advantage of this example to illustrate how Definition~\ref{defn_16} is put to use to represent rules. Recall that $\bigwedge\varnothing$ is valid and $\bigvee\varnothing$ is invalid. For the propositional rules,
\begin{itemize}
\item
$\varphi^+_{q_1}$ and $\varphi^-_{q_1}$ represent the fact $q_1$,
\item
$\varphi^+_{q_2}$ and $\varphi^-_{q_2}$ represent the rules $q_2\leftarrow q_3$ and $\neg q_2\leftarrow\neg q_3$,
\item
$\varphi^+_{q_3}$ and $\varphi^-_{q_3}$ represent the rules $q_3\leftarrow q_2$ and $\neg q_3\leftarrow\neg q_2$,
\item
$\varphi^+_{q_4}$ and $\varphi^-_{q_4}$ represent the rule $q_4\leftarrow q_4$, and
\item
$\varphi^+_{q_5}$ and $\varphi^-_{q_5}$ represent the rule $q_5\leftarrow\neg q_5$.
\end{itemize}
Now to the first-order rules.
\begin{itemize}
\item
The formulas $\varphi^+_{p_i}$, $i\in\{1,2\}$, represent the rule
\begin{align*}
&p_i(v_1)\leftarrow v_1\doteq\NUM{0}\vee\exists v_0\bigl(v_1\doteq s(s(v_0))\wedge p_i(v_0)\bigr) \\
\intertext{which could be rewritten as the following fact and rule.}
&p_i(\NUM{0}) \\
&p_i(s(s(v_1)))\leftarrow p_i(v_1)
\end{align*}
So for $i\in\{1,2\}$, $\varphi^+_{p_i}$ allows one to generate all literals of the form $p_i(\NUM{2n})$, $n\in\NSET$. It is easily verified that for $i\in\{1,2\}$, $\varphi^-_{p_i}$ allows one to generate all literals of the form $\neg p_i(\NUM{2n+1})$, $n\in\NSET$. More precisely, the rule represented by $\varphi_{p_1}^-$, namely
\[
\neg p_1(v_1)\leftarrow v_1\not\doteq\NUM{0}\wedge\forall v_0\bigl(v_1\not\doteq s(s(v_0))\vee\neg p_1(v_0)\bigr)
\]
could be naturally implemented from $\{p_1(\NUM{0}),p_1(s(s(v_1)))\leftarrow p_1(v_1)\}$ using negation as finite failure, and its syntax is naturally related to Cark's completion of the set $\{p_1(\NUM{0}),\ p_1(s(s(v_1)))\leftarrow p_1(v_1)\}$.
\item
The rule $\neg p_2(v_1)\leftarrow\varphi^-_{p_2}$, namely
\begin{align*}
&\neg p_2(v_1)\leftarrow v_1\doteq s(\NUM{0})\vee\exists v_0\bigl(v_1\doteq s(s(v_0))\wedge\neg p_2(v_0)\bigr) \\
\intertext{is very similar to the rule $p_2(v_1)\leftarrow\varphi^+_{p_2}$, and could be rewritten}
&\neg p_2(s(\NUM{0})) \\
&\neg p_2(s(s(v_1)))\leftarrow\neg p_2(v_1)
\end{align*}
to generate $\{\neg p_2(\NUM{2n+1})\mid n\in\NSET\}$ similarly to the way $\{p_2(\NUM{2n})\mid n\in\NSET\}$ would be generated using $p_2(v_1)\leftarrow\varphi^+_{p_2}$.
\item
The formulas $\varphi^+_{p_3}$ and $\varphi^-_{p_3}$ offer a third way of generating the set of even numbers and its complement, with both the positive rule $p_3(v_1)\leftarrow\varphi^+_{p_3}$ and the negative rule $\neg p_3(v_1)\leftarrow\varphi^-_{p_3}$ being used alternatively, starting with the positive rule.
\item
Finally, $\varphi^+_{p_4}$ and $\varphi^-_{p_4}$ represent the rules
\begin{align*}
p_4(v_1)&\leftarrow p_4(s(s(v_1))) \\
\neg p_4(v_1)&\leftarrow\neg p_4(s(s(v_1)))
\end{align*}
and would not generate any literal.
\end{itemize}

The second item in Definition~\ref{defn_8} captures the unique name axioms that come with the definition of Clark's completion of a logic program. Making \EQ\ a logical symbol amounts to building into the logic a notion of identity stronger than the usual, less restrictive notion of equality, that in our framework is nonlogical and has to be axiomatised if needed: $=$ is then put into \VOCA\ and its intended interpretation captured by any formal logic program $(\varphi_\wp^\epsilon)_{(\wp,\epsilon)\in\IND}$ such that $\varphi_=^+$ is of the form $\bigvee X$ where $X$ is a superset of
\begin{multline*}
\Bigl\{v_1\doteq v_2,\;v_2=v_1,\;\exists v_0(v_1=v_0\wedge v_0= v_2)\Bigr\}\;\cup\\
\Bigl\{\exists v_3\dots\exists v_{3+2n-1}\bigl(v_3=v_{3+n}\wedge\dotsb\wedge v_{3+n-1}=v_{3+2n-1}\,\wedge\\
v_1\doteq f(v_3,\dotsc,v_{3+n-1})\wedge v_2\doteq f(v_{3+n},\dotsc,v_{3+2n-1})\bigr)\Bigm|\\
n\in\NSET,\mbox{ $f$ is an $n$-ary function symbol in \VOCA}\Bigr\},
\end{multline*}
$\varphi_=^-$ is of the form $\bigvee X$ where $X$ contains
\[
\exists v_0\bigl((v_0= v_1\wedge v_0\neq v_2)\vee(v_0\neq v_1\wedge v_0=v_2)\bigr)
\]
and for all $n\in\NSET$, $\wp\in\PRDVN{n}$ and $\epsilon\in\{+,-\}$, $\varphi_\wp^\epsilon$ is of the form $\bigvee X$ where $X$ contains
\[
\exists v_{n+1}\dots\exists v_{2n}\bigl(v_1=v_{n+1}\wedge\dotsb\wedge v_n=v_{2n}\wedge\wp^{\epsilon}(v_{n+1},\dotsc,v_{2n})\bigr).
\]
Identity is a key notion in logic programming as it is at the heart of the unification algorithm, and the usual approach is to treat identity and equality as equivalent, with the restriction to the class of Herbrand interpretations as a justification for the identification of both notions. Our approach consists in logically defining identity from \VOCAB, the vocabulary used to describe a structure, and in axiomatising equality from \VOCA, the vocabulary used to talk about a structure. As a consequence, equality  and Herbrand structures are not interdependent: if infinitely many closed terms are `unspeakable of' then equality as axiomatised above behaves equivalently to the way it behaves w.r.t.\ the classical notion of logical consequence. 

\begin{defn}\label{defn_18}
Given a formal logic program $\FLP=(\varphi_\wp^\epsilon)_{(\wp,\epsilon)\in\IND}$, the \emph{classical logical form of \FLP} is defined as
\[
\bigl\{\varphi_\wp^\epsilon\rightarrow\wp^\epsilon(v_1,\dotsc,v_n)\bigm| n\in\NSET,\,\wp\in\PRDVN{n},\,\epsilon\in\{+,-\}\bigr\}.
\]
\end{defn}

Of course, $\WMODELS$ applied to the classical logical form of a formal logic program \FLP\ does not adequately capture the logical meaning of \FLP. An appropriate logical reading of a formal logic program, which amounts to an appropriate denotational semantics, requires more than reading the arrow that links the left hand side and right hand side of a rule as a logical implication: it requires the explicit use of a modal operator of necessity to capture the notion of derivability, or provability, in the style of epistemic logic~\cite{Moore:1985,Marek:1991}. We will complete this task in another paper.

\begin{nota}\label{nota_19}
Given a formal logic program \FLP, we let $\CLF(\FLP)$ denote the classical logical form of \FLP.
\end{nota}

The general logic programs that are the object of Kripke-Kleene semantics, the well-founded semantics and the stable model semantics can be seen as a particular case of formal logic programs where the negative rules are fully determined by the positive rules and can be left implicit; they are in one-to-one correspondence with the formal logic programs defined next.

\begin{defn}\label{defn_20}
Let a formal logic program $\FLP=(\varphi_\wp^\epsilon)_{(\wp,\epsilon)\in\IND}$ be given.
We say that \FLP\ is \emph{symmetric} iff for all $\wp\in\PRDV$,
$\varphi_\wp^-$ is identical to $\SIM\varphi_\wp^+$.
\end{defn}

The next definition introduces a notion that is a key property of symmetric formal logic programs.

\begin{defn}\label{defn_21}
We say that a formal logic program $(\varphi_\wp^\epsilon)_{(\wp,\epsilon)\in\IND}$ is \emph{locally consistent} iff for all $\wp\in\PRDV$, no closed instance of
$\varphi_\wp^+\wedge\varphi_\wp^-$ has a model in \WORLDS.
\end{defn}

\begin{property}\label{property_22}
A symmetric formal logic program is locally consistent. 
\end{property}

\subsection{Generated literals}

The mechanistic view on (the rules of) a formal logic program \FLP\ presented in Section~\ref{mvr} allows one to talk about the literals over \VOCA\ generated by \FLP; these literals make up a set that we denote by $\LB\FLP\RB$. More precisely, a literal $\psi$ over \VOCA\ is generated by \FLP\ and put into $\LB\FLP\RB$ if it is possible to successively fire rules, starting with rules whose body can be unconditionally activated (such as $\bigwedge\varnothing$), till enough literals have been generated and put into $\LB\FLP\RB$ so that there exists a rule in \FLP\ of the form $\chi\leftarrow\xi$ and a substitution $\theta$ such that $\psi$ is $\chi\theta$ and for all closed substitutions $\theta'$, $\xi(\theta\theta')$ can be activated thanks to the literals in $\LB\FLP\RB$; the notation that follows will allow us to easily refer to a formula (determined by $\psi$ and \FLP) of the form $\xi\theta''$ where $\theta''$ is $\theta$ with some of the variables in its range being possibly renamed so that none of the variables that occurs in $\psi$ (that is, in $\chi\theta$) is captured when applying the substitution $\theta''$ to $\xi$.

\begin{nota}\label{nota_23}
Let a formal logic program \FLP\ be given. For all $n\in\NSET$, $\wp\in\PRDVN{n}$, terms $t_1$, \dots, $t_n$ and $\epsilon\in\{+,-\}$, we let $\FLP[\wp^\epsilon(t_1,\dots,t_n)]$ denote a formula of the form $\varphi_\wp^\epsilon[t'_1/v_1,\dots,t'_n/v_n]$ whose closed instances are precisely the formulas of the form $\varphi_\wp^\epsilon[t''_1/v_1,\dots,t''_n/v_n]$ with $(t''_1,\dotsc,t''_n)$ any closed instance of $(t_1,\dotsc,t_n)$.
\end{nota}

In the context of Notation~\ref{nota_23}, observe that if none of the variables occurring in one of $t_1$, \dots, $t_n$ is captured by quantifiers in $\varphi_\wp^\epsilon$ when simultaneously substituting $v_1$, \dots, $v_n$ in $\varphi_\wp^\epsilon$ by $t_1$, \dots, $t_n$, respectively, then a natural choice for $\FLP[\wp^\epsilon(t_1,\dots,t_n)]$ is $\varphi_\wp^\epsilon[t_1/v_1,\dotsc,t_n/v_n]$ itself.
Using Notation~\ref{nota_23}, one can then concisely define the set of literals over \VOCA\ generated by a formal logic program as a fixed point:

\begin{nota}\label{nota_24}
Given a formal logic program \FLP, we denote by $\LB\FLP\RB$ the (unique) $\subseteq$-minimal set of literals over \VOCA\ that forces $\FLP[\psi]$ for all $\psi\in\LB\FLP\RB$.
\end{nota}

Of course, we could alternatively define $\LB\FLP\RB$ in terms of a transfinite construction and collect at some round, indexed by an ordinal, the set of literals over \VOCA\ that can be generated from \FLP\ by activating some instances of the bodies of some of \FLP's rules thanks to the literals generated at previous rounds. This construction is defined in Notation~\ref{nota_25}, and the fact that it is an alternative definition to  $\LB\FLP\RB$ is stated as Property~\ref{property_26}.

\begin{nota}\label{nota_25}
Let a formal logic program \FLP\ be given. Inductively define a sequence $(\LB\FLP\RB_\alpha)_{\alpha\in\ORD}$ of sets of literals over \VOCA\ as follows. Let an ordinal $\alpha$ be given and assume that for all $\beta<\alpha$, $\LB\FLP\RB_\beta$ has been defined. Then denote by $\LB\FLP\RB_\alpha$ the set of all literals $\psi$ over \VOCA\ with $\bigcup_{\beta<\alpha}\LB\FLP\RB_\beta\FORCES\FLP[\psi]$.
\end{nota}

\begin{property}\label{property_26}
For all formal logic programs \FLP, $\LB\FLP\RB=\bigcup_{\alpha\in\ORD}\LB\FLP\RB_\alpha$.
\end{property}

\begin{ex}\label{ex_27}
If \FLP\ is the formal logic program of Example~\ref{ex_17} then
\[
\LB\FLP\RB=\bigl\{p_i(\NUM{2n}),\,\neg p_i(\NUM{2n+1})\mid i\in\{1,2,3\},\,n\in\NSET\bigr\}\cup\{q_1\}.
\]
\end{ex}

It is easy to verify that the set of literals over \VOCA\ generated by a formal logic program is closed under forcing.

\begin{property}\label{property_28}
For all formal logic programs \FLP\ and literals $\psi$ over \VOCA,
$\LB\FLP\RB\FORCES\psi$ iff $\psi\in\LB\FLP\RB$.  
\end{property}

Local consistency as introduced in Definition~\ref{defn_21} will play a pivotal role in the statements of some propositions, but the more general notion of plain consistency given next is a better counterpart to the classical concept of a consistent theory.

\begin{defn}\label{defn_29}
A formal logic program \FLP\ is said to be \emph{consistent} just in case $\LB\FLP\RB$ is consistent.
\end{defn}

\begin{property}\label{property_30}
Every locally consistent formal logic program is consistent.
\end{property}

Let a formal logic program $\FLP=(\varphi_\wp^\epsilon)_{(\wp,\epsilon)\in\IND}$ be given. When $\VOCA=\VOCAB$, the definition of $\LB\FLP\RB$ can involve closed literals only---a consequence of Property~\ref{property_28} and the next property. In the general case, $\LB\FLP\RB$ is a set of possibly nonclosed literals, and some rules might fire because their bodies are activated thanks to such literals. For instance, assume that \VOCA\ contains a unary predicate symbol $p$ and a nullary predicate symbol $q$, $\varphi_p^+=\bigwedge\varnothing$, and $\varphi_q^+=\forall v_0\,p(v_0)$. Then $\LB\FLP\RB$ contains $p(v_0)$, hence it contains $q$. Also, $\LB\FLP\RB$ contains $p(t)$ for all terms $t$ over \VOCA, hence in particular for all closed terms $t$ over \VOCA. Still, if at least one of \VOCAB's constants does not belong to \VOCA, then the set of all closed members of $\LB\FLP\RB$ of the form $p(t)$ (with $t$ a closed term over \VOCA) does not force $\forall v_0\,p(v_0)$, which shows that the next property would not hold if the assumption $\VOCA=\VOCAB$ was dropped.

\begin{property}\label{property_31}
Let a formal logic program \FLP\ be given. If $\VOCA=\VOCAB$ then the set of closed members of $\LB\FLP\RB$ is the (unique) $\subseteq$-minimal set $X$ of closed literals with $X\FORCES\{\FLP[\psi]\mid\psi\in X\}$.
\end{property}

The classical logical form of a formal logic program \FLP, formalised in Definition~\ref{defn_18}, does not capture in a satisfactory way the logical meaning of \FLP, but it is still well behaved, as expressed by the property and the corollary that follow.

\begin{property}\label{property_32}
For all formal logic programs \FLP, $\CLF(\FLP)\WMODELS\LB\FLP\RB$.
\end{property}

\begin{cor}\label{cor_33}
For all formal logic programs \FLP, if $\LB\FLP\RB$ is complete then the set of closed instances of atoms in $\LB\FLP\RB$ is a model of $\CLF(\FLP)$.
\end{cor}

\subsection{Characterisation of Kripke-Kleene semantics}

The characterisation is based on the following definition.

\begin{defn}\label{defn_34}
A \emph{partial interpretation} (\emph{over\/ \VOCAB}) is a consistent set of closed literals.
\end{defn}

Kripke-Kleene semantics is usually presented in a 3-valued logical setting.
The relationship between Definition~\ref{defn_34} and a 3-valued logical setting is the following. Let $M$ be a partial interpretation, and let a closed atom $\varphi$ be given. Then the truth value of $\varphi$ in $M$ can be set to \true\ if $\varphi\in M$, to \false\ if $\neg\varphi\in M$, and to a third value or to `undefined' otherwise.
Definition~\ref{defn_35} then generalises the notion of a partial model of a general logic program---that as we have pointed out, can be seen as a symmetric formal logic program whose negative rules have not been explicitly written.

\begin{defn}\label{defn_35}
Let a formal logic program $\FLP=(\varphi_\wp^\epsilon)_{(\wp,\epsilon)\in\IND}$ be given. A \emph{partial model of\/ \FLP} is a partial interpretation $M$ such that for all $n\in\NSET$, $\wp\in\PRDVN{n}$, closed terms $t_1$, \dots, $t_n$ and $\epsilon\in\{+,-\}$, $M$ contains $\wp^\epsilon(t_1,\dotsc,t_n)$ iff $M$ forces $\varphi_\wp^\epsilon[t_1/v_1,\dotsc,t_n/v_n]$.
\end{defn}

Given a formal logic program \FLP, a \emph{$\subseteq$-minimal partial model of\/ \FLP} is referred to more simply as a minimal partial model of \FLP.
Proposition~\ref{prop_36} expresses that the generalisation of Kripke-Kleene semantics given in Definition~\ref{defn_35} is equivalent to our base semantics of a consistent formal logic program, provided that \VOCA\ is equal to \VOCAB, which is the underlying assumption of all frameworks where that semantics is considered. Note that Proposition~\ref{prop_36} still applies to more general frameworks as it deals with formal logic programs that might not be symmetric.

\begin{prop}\label{prop_36}
Assume that $\VOCA=\VOCAB$. Let a consistent formal logic program \FLP\ be given. Then \FLP\ has a unique minimal partial model, which is nothing but the set of closed instances of members of $\LB\FLP\RB$.
\end{prop}

\begin{proof}
Let $X$ denote the set of partial models of \FLP.
It is immediately verified that:
\begin{itemize}
\item
the set of closed instances of members of $\LB\FLP\RB$ is included in $\bigcap X$;
\item
the set of closed instances of members of $\LB\FLP\RB$ belongs to $X$.
\end{itemize}
Hence $\bigcap X$, being equal to the set of closed instances of members of $\LB\FLP\RB$, is a partial model of \FLP.
\end{proof}

\section{Extensors, and relationships to particular semantics}

\subsection{Extensors}\label{uuee}

We now formalise the operation, discussed in Section~\ref{mch}, of transforming a formal logic program \FLP\ into another formal logic program $\FLP\PLUS E$, where $\Omega$  selects some occurrences of literals in the bodies of \FLP's rules and $E$ is a set of literals, the intended meaning of $\FLP\PLUS E$ being: ``in \FLP, assume $E$ in the contexts indicated by $\Omega$.''
Definition~\ref{defn_37} defines the kind of formal object denoted by $\Omega$. Notation~\ref{nota_38} specifies three particular cases the first two of which will play a special role in relation to the stable model and the well-founded semantics. Recall Definition~\ref{defn_6} for the notion of an occurrence of a literal in a formula.

\begin{defn}\label{defn_37}
Let a formal logic program $\FLP=(\varphi_\wp^\epsilon)_{(\wp,\epsilon)\in\IND}$ be given.
A \emph{literal marker for\/ \FLP} is a sequence of the form
$(O_\wp^\epsilon)_{(\wp,\epsilon)\in\IND}$ where for all $\wp\in\PRDV$ and $\epsilon\in\{+,-\}$, $O_\wp^\epsilon$ is a set of occurrences of literals in $\varphi_\wp^\epsilon$.
\end{defn}

\begin{nota}\label{nota_38}
Let a formal logic program \FLP\ and a literal marker $\Omega$ for \FLP\ be given.
Write \FLP\ as $(\varphi_\wp^\epsilon)_{(\wp,\epsilon)\in\IND}$ and $\Omega$ as $(O_\wp^\epsilon)_{(\wp,\epsilon)\in\IND}$.
\begin{itemize}
\item
If for all $\wp\in\PRDV$, $O_\wp^+$ is empty and $O_\wp^-$ is the set of all occurrences of negated atoms in $\varphi_\wp^-$, then we denote $\Omega$ by \SPE{\cdot,-}.
\item
If for all $\wp\in\PRDV$ and $\epsilon\in\{+,-\}$, $O_\wp^\epsilon$ is the set of all occurrences of negated atoms in $\varphi_\wp^\epsilon$, then we denote $\Omega$ by \SPE{-,-}.
\item
If for all $\wp\in\PRDV$ and $\epsilon\in\{+,-\}$, $O_\wp^\epsilon$ is the set of all occurrences of literals in $\varphi_\wp^\epsilon$, then we denote $\Omega$ by \SPE{\pm,\pm}.
\end{itemize}
\end{nota}

In Section~\ref{mch}, we gave the following introductory example. Assume that \VOCA\ contains the constant \NUM{0}, the unary function symbol $s$ and three unary predicate symbols $p$, $q$ and $r$. Let \FLP\ be a formal logic program, say $(\varphi_\wp^\epsilon)_{(\wp,\epsilon)\in\IND}$, such that $\varphi_p^+$ is equal to
\[
\bigl(p(v_1)\vee q(v_1)\bigr)\wedge\bigl(p(v_1)\vee r(v_1)\bigr).
\]
Let $\Omega=(O_\wp^\epsilon)_{(\wp,\epsilon)\in\IND}$ be the literal marker for \FLP\ such that $O_p^+$ is equal to
\[
\bigl\{\{\varphi_p^+,\,p(v_1)\vee q(v_1),\,q(v_1)\},\,\{\varphi_p^+,\,p(v_1)\vee r(v_1),\,p(v_1)\}\bigr\}
\]
which corresponds to marking $\varphi_p^+$ as
\[
\bigl(p(v_1)\vee \underset{\checkmark}{q(v_1)}\bigr)\wedge\bigl(\underset{\checkmark}{p(v_1)}\vee\ r(v_1)\bigr).
\]
Let $E$ be defined as $\bigl\{p(\NUM{2n})\mid n\in\NSET\bigr\}$. Then $\FLP\PLUS E$ is a formal logic program, say $(\psi_\wp^\epsilon)_{(\wp,\epsilon)\in\IND}$, such that $\psi_p^+$ will be defined in such a way that it is logically equivalent in \WORLDS\ to
\[
\Bigl(p(v_1)\vee q(v_1)\Bigr)\wedge\Bigl(p(v_1)\vee\bigvee_{n\in\NSET}v_1\doteq\NUM{2n}\vee r(v_1)\Bigr).
\]
If we modify the example and assume that $E$ is rather set to $\{p(s(s(v_0)))\}$, then $\psi_p^+$ will be defined in such a way that it is logically equivalent in \WORLDS\ to
\[
\Bigl(p(v_1)\vee q(v_1)\Bigr)\wedge\Bigl(p(v_1)\vee\exists v_0\bigl(v_1\doteq s(s(v_0))\bigr)\vee r(v_1)\Bigr).
\]
The eventual definition of $\FLP\PLUS E$ for arbitrary choices of \FLP, $\Omega$ and $E$, will be a straightforward generalisation of those examples. One should keep in mind that $E$ will eventually be chosen as a set of literals over \VOCA\ (as opposed to a set of literals over \VOCAB), with \VOCAB\ and \VOCA\ being possibly different, which implies that again, we cannot assume in full generality that $E$ can be restricted to consist of closed literals only.

The notation that follows should be thought of as recording the set of all possible substitutions thanks to which a given formula $\varphi$ can be shown to subsume some member of a set $E$ of formulas.
  
\begin{nota}\label{nota_39}
Given a formula $\varphi$, $n\in\NSET$, distinct variables $x_1$, \dots, $x_n$ with $\VAR{\varphi}=\{x_1,\dotsc,x_n\}$, and a set $E$ of formulas, we let \UNIF{\varphi}{E}
denote the set of all formulas of the form
\[
\exists y_1\dots\exists y_m(x_1\EQ t_1\wedge\dotsc\wedge x_n\EQ t_n)
\]
where $t_1$, \dots, $t_n$ are terms over \VOCA, $m$ is a member of \NSET, $y_1$, \dots, $y_m$ are distinct variables, all distinct from $x_1$, \dots, $x_n$, $\{y_1,\dots,y_m\}$ is the set of variables that occur in at least one of $t_1$, \dots, $t_n$ and for all closed terms $t'_1$, \dots, $t'_n$, if $(t'_1,\dots,t'_n)$ is an instance of $(t_1,\dotsc,t_n)$ then $\varphi[t'_1/x_1,\dotsc,t'_n/x_n]$ is an instance of a member of $E$.
\end{nota}

In the propositional case, $n=0$ and Notation~\ref{nota_39} simplifies the definition of $\UNIF{\varphi}{E}$ to $\{\bigwedge\varnothing\}$ if $\varphi\in E$, and to $\varnothing$ otherwise.
The next notation will describe the operations of strengthening or weakening some occurrences of literals in a formula: given a formula $\varphi$, a set $O$ of occurrences of literals in $\varphi$ and a set $E$ of literals,
\begin{itemize}
\item
\selects{E}{O}{\varphi} will be a formula obtained from $\varphi$ by assuming that every occurrence of a literal in $\varphi$ that belongs to $O$ is false unless it subsumes some member of $E$;
\item
\extends{E}{O}{\varphi} will be a formula obtained from $\varphi$ by assuming that every occurrence of a literal in $\varphi$ that belongs to $O$ is true if it subsumes some member of $E$.
\end{itemize}
Note that \selects{E}{O}{\varphi} and \extends{E}{O}{\varphi} do not denote two notions that differ only in that one refers to ``false'' when the other refers to ``true.'' This is maybe more easily observed thanks to the following alternative, but less precise, informal description of \selects{E}{O}{\varphi} and \extends{E}{O}{\varphi}: an occurrence of a literal in $\varphi$ that belongs to $O$ is true in \selects{E}{O}{\varphi} iff it subsumes some member of $E$, whereas an occurrence of a literal in $\varphi$ that belongs to $O$ is true in \extends{E}{O}{\varphi} iff it subsumes a member of $E$ or if it can be shown to be true.

The first operation prepares the technical definition of a formal logic program obtained from \FLP\ and $E$, and denoted $\FLP\SELECT E$, that will be useful for formalising in our setting the answer-set and the stable model semantics. The second operation prepares the definition of $\FLP\PLUS E$. Hence to develop our framework, only the notation \extends{E}{O}{\varphi} is needed: the notation \selects{E}{O}{\varphi} is used only to reformulate the usual definitions of the answer-set and the stable model semantics in a way that will make it easier to establish their relationship to our setting. To relate our framework to the stable model semantics, $\Omega$ will be set to \SPE{-,-}, and to relate it to the well-founded semantics, $\Omega$ will be set to either \SPE{-,-} or \SPE{\cdot,-} (both options are equally suitable), which prompts for a special notation, that of Notation~\ref{nota_41}. 
Both $\FLP\SELECT E$ and $\FLP\PLUS E$ are formally defined in Notation~\ref{nota_43}.

\begin{nota}\label{nota_40}
Let $E$ be a set of literals.
We inductively define for all formulas $\varphi$ and sets $O$ of occurrences of literals in $\varphi$ two formulas \selects{E}{O}{\varphi} and \extends{E}{O}{\varphi}. Let $\varphi\in\INFLVOCAB$ and a set $O$ of occurrences of literals in $\varphi$ be given. 
\begin{itemize}
\item
Suppose that $\varphi$ is an identity, a distinction or a literal.
\begin{itemize}
\item
If $O=\varnothing$ then both \selects{E}{O}{\varphi} and \extends{E}{O}{\varphi} are $\varphi$.
\item
If $O=\bigl\{\{\varphi\}\bigr\}$ then \selects{E}{O}{\varphi} is $\bigvee\UNIF{\varphi}{E}$ and \extends{E}{O}{\varphi} is $\bigvee\{\varphi\}\cup\UNIF{\varphi}{E}$.
\end{itemize}
\item
Suppose that $\varphi$ is of the form $\bigvee X$ or $\bigwedge X$. For all $\psi\in X$, let
$O_\psi$ be the (unique) set of occurrences of literals in $\psi$, say $o$, with $o\cup\{\varphi\}\in O$.
\begin{itemize}
\item
If $\varphi$ is the formula $\bigvee X$ then \selects{E}{O}{\varphi} is $\bigvee\{\selects{E}{O_\psi}{\psi}\mid\psi\in X\}$ and \extends{E}{O}{\varphi} is $\bigvee\{\extends{E}{O_\psi}{\psi}\mid\psi\in X\}$.
\item
If $\varphi$ is the formula $\bigwedge X$ then \selects{E}{O}{\varphi} is $\bigwedge\{\selects{E}{O_\psi}{\psi}\mid\psi\in X\}$ and \extends{E}{O}{\varphi} is $\bigwedge\{\extends{E}{O_\psi}{\psi}\mid\psi\in X\}$.
\end{itemize}
\item
Suppose that $\varphi$ is of the form $\exists x\,\psi$ or $\forall x\,\psi$.
Let $O_\psi$ be the (unique) set of occurrences of literals in $\psi$, say $o$, with $o\cup\{\varphi\}\in O$.
\begin{itemize}
\item
If $\varphi$ is the formula $\exists x\,\psi$ then \selects{E}{O}{\varphi} and \extends{E}{O}{\varphi} are the existential closure of \selects{E}{O_\psi}{\psi} and $\exists x\extends{E}{O_\psi}{\psi}$, respectively.\footnote{We cannot write $\exists x\selects{E}{O_\psi}{\psi}$ as $x$ might not occur free in \selects{E}{O_\psi}{\psi}.}
\item
If $\varphi$ is the formula $\forall x\,\psi$ then \selects{E}{O}{\varphi} and \extends{E}{O}{\varphi} are the universal closure of \selects{E}{O_\psi}{\psi} and $\forall x\extends{E}{O_\psi}{\psi}$, respectively.\footnote{Similarly, we cannot write $\forall x\selects{E}{O_\psi}{\psi}$ as $x$ might not occur free in \selects{E}{O_\psi}{\psi}.}
\end{itemize}
\end{itemize}
\end{nota}

\begin{nota}\label{nota_41}
Given $\varphi\in\INFLVOCAB$ and a set $E$ of literals, and letting $O$ be the set of occurrences of negated atoms in $\varphi$, we write \selects{E}{-}{\varphi} for \selects{E}{O}{\varphi} and \extends{E}{-}{\varphi} for \extends{E}{O}{\varphi}.
\end{nota}

\begin{ex}\label{ex_42}
Suppose that \FLP\ is the formal logic program of Example~\ref{ex_17} and
\[
E=\{p_3(\NUM{2}),\,p_3(\NUM{3}),\,\neg p_3(\NUM{1}),\,\neg p_3(\NUM{2}),\,p_4(\NUM{2}),\,\neg p_4(\NUM{1}),\,\neg q_5\}.
\]
\begin{itemize}
\item
As \UNIF{\neg p_3(v_0)}{E} is $\{v_0\EQ\NUM{1},v_0\EQ\NUM{2}\}$, \selects{E}{-}{\varphi_{p_3}^+} and \extends{E}{-}{\varphi_{p_3}^+} are
\[
v_1\EQ\NUM{0}\vee\exists v_0\bigl( v_1\EQ s(v_0)\wedge(v_0\EQ\NUM{1}\vee v_0\EQ\NUM{2})\bigr),
\]
and
\[
v_1\EQ\NUM{0}\vee\exists v_0\bigl( v_1\EQ s(v_0)\wedge(\neg p_3(v_0)\vee v_0\EQ\NUM{1}\vee v_0\EQ\NUM{2})\bigr),
\]
which are logically equivalent in \WORLDS\ to
\[
v_1\EQ\NUM{0}\vee v_1\EQ\NUM{2}\vee v_1\EQ\NUM{3}
\]
and
\[
v_1\EQ\NUM{0}\vee v_1\EQ\NUM{2}\vee v_1\EQ\NUM{3}\vee\exists v_0\bigl(v_1\EQ s(v_0)\wedge\neg p_3(v_0)\bigr),
\]
respectively.
\item
As $\varphi_{p_4}^+$ does not contain any occurrence of a negated atom, \selects{E}{-}{\varphi_{p_4}^+} and \extends{E}{-}{\varphi_{p_4}^+} are both identical to $\varphi_{p_4}^+$.
\item
As $\neg p_4(s(s(v_1)))$ does not unify with $\neg p_4(\NUM{1})$ and $\neg q_3$ does not belong to $E$, \selects{E}{-}{\varphi_{p_4}^-} and \selects{E}{-}{\varphi_{q_2}^-} are both identical to $\bigvee\varnothing$, while \extends{E}{-}{\varphi_{p_4}^-} and \extends{E}{-}{\varphi_{q_2}^-} are logically equivalent in \WORLDS\ to $\varphi_{p_4}^-$ and $\varphi_{q_2}^-$, respectively.
\item
As $\neg q_5$ belongs to $E$, \UNIF{\neg q_5}{E} is $\{\bigwedge\varnothing\}$, and \selects{E}{-}{\varphi_{q_5}^+} and \extends{E}{-}{\varphi_{q_5}^+} are both logically equivalent in \WORLDS\ to $\bigwedge\varnothing$.
\end{itemize}
\end{ex}

\begin{nota}\label{nota_43}
Let a formal logic program \FLP\ and a literal marker $\Omega$ for \FLP\ be given. Write \FLP\ as $(\varphi_\wp^\epsilon)_{(\wp,\epsilon)\in\IND}$ and
$\Omega$ as $(O_\wp^\epsilon)_{(\wp,\epsilon)\in\IND}$. Let a set $E$ of 
literals be given.
\begin{itemize}
\item
We let $\FLP\SELECT E$ denote
$(\selects{E}{O_\wp^\epsilon}{\varphi_\wp^\epsilon})_{(\wp,\epsilon)\in\IND}$.
\item
We let $\FLP\PLUS E$ denote
$(\extends{E}{O_\wp^\epsilon}{\varphi_\wp^\epsilon})_{(\wp,\epsilon)\in\IND}$.
\end{itemize}
\end{nota}

\begin{property}\label{property_44}
For all formal logic programs \FLP, literal markers $\Omega$ for \FLP\ and
sets $E$ of literals, $\LB\FLP\RB\subseteq\LB\FLP\PLUS E\RB$.
\end{property}

The next property will be applied in the particular case where one of $E$ and $F$ denotes a set of literals over \VOCA, and the other the set of its closed instances.

\begin{property}\label{property_45}
Let a formal logic program \FLP, a literal marker $\Omega$ for \FLP, and
two sets $E$ and $F$ of literals be such that $E$ and $F$ have the same closed instances. Then $\LB\FLP\SELECT E\RB$ is equal to $\LB\FLP\SELECT F\RB$ and $\LB\FLP\PLUS E\RB$ is equal to $\LB\FLP\PLUS F\RB$.
\end{property}

The next lemma will play a crucial role in relating our framework to the answer-set and the stable model semantics.

\begin{lem}\label{lem_46}
Let a formal logic program \FLP, a literal marker $\Omega$ for \FLP, and a set $E$ of literals be given. Then $\LB\FLP\SELECT E\RB\subseteq\LB\FLP\PLUS E\RB$. Also, if all closed instances of members of $\LB\FLP\SELECT E\RB$ are instances of members of $E$ then $\LB\FLP\PLUS E\RB=\LB\FLP\SELECT E\RB$.
\end{lem}

\begin{proof}
Write $\Omega=(O_\wp^\epsilon)_{(\wp,\epsilon)\in\IND}$.

Let us verify the first part of the lemma. By Property~\ref{property_26}, it suffices to show that for all ordinals $\alpha$, $\LB\FLP\SELECT E\RB_\alpha$ is a subset of $\LB\FLP\PLUS E\RB_\alpha$. Proof is by induction. Let an ordinal $\alpha$ be given, and assume that for all $\beta<\alpha$, $\LB\FLP\SELECT E\RB_\beta\subseteq\LB\FLP\PLUS E\RB_\beta$. Let $n\in\NSET$, $\wp\in\PRDVN{n}$, terms $t_1$, \dots, $t_n$ over \VOCA\ and $\epsilon\in\{+,-\}$ be such that $\wp^\epsilon(t_1,\dotsc,t_n)\in\LB\FLP\SELECT E\RB_\alpha$. Then $\bigcup_{\beta<\alpha}\LB\FLP\SELECT E\RB_\beta\FORCES\selects{E}{O_\wp^\epsilon}{\FLP\bigl[\wp^\epsilon[t_1/v_1,\dotsc,t_n/v_n]\bigr]}$. Moreover, it is immediately verified that $\bigl\{\selects{E}{O_\wp^\epsilon}{\FLP\bigl[\wp^\epsilon[t_1/v_1,\dotsc,t_n/v_n]\bigr]}
\bigr\}$ logically implies $\extends{E}{O_\wp^\epsilon}{\FLP\bigl[\wp^\epsilon[t_1/v_1,\dotsc,t_n/v_n]\bigr]}$ in \WORLDS. This together with the inductive hypothesis implies that $\bigcup_{\beta<\alpha}\LB\FLP\PLUS E\RB_\beta$ forces $\extends{E}{O_\wp^\epsilon}{\FLP\bigl[\wp^\epsilon[t_1/v_1,\dotsc,t_n/v_n]\bigr]}$; hence $\wp^\epsilon(t_1,\dotsc,t_n)\in\LB\FLP\PLUS E\RB_\alpha$, completing the verification that $\LB\FLP\SELECT E\RB\subseteq\LB\FLP\PLUS E\RB$.

Assume that all closed instances of members of $\LB\FLP\SELECT E\RB$ are instances of members of $E$. Suppose for a contradiction that $\LB\FLP\PLUS E\RB\nsubseteq\LB\FLP\SELECT E\RB$. By Property~\ref{property_26}, let ordinal $\alpha$ be least with $\LB\FLP\PLUS E\RB_\alpha\nsubseteq\LB\FLP\SELECT E\RB$. Let $n\in\NSET$, $\wp\in\PRDVN{n}$, terms $t_1$, \dots, $t_n$ over \VOCA\ and $\epsilon\in\{+,-\}$ be such that $\wp^\epsilon(t_1,\dotsc,t_n)$ belongs to $\LB\FLP\PLUS E\RB_\alpha\setminus\LB\FLP\SELECT E\RB$. Then $\bigcup_{\beta<\alpha}\LB\FLP\PLUS E\RB_\beta\FORCES\extends{E}{O_\wp^\epsilon}{\FLP\bigl[\wp^\epsilon[t_1/v_1,\dotsc,t_n/v_n]\bigr]}$; so by the choice of $\alpha$, $\LB\FLP\SELECT E\RB\FORCES\extends{E}{O_\wp^\epsilon}{\FLP\bigl[\wp^\epsilon[t_1/v_1,\dotsc,t_n/v_n]\bigr]}$, which together with the assumption on $E$, easily implies that $\LB\FLP\SELECT E\RB$ forces $\selects{E}{O_\wp^\epsilon}{\FLP\bigl[\wp^\epsilon[t_1/v_1,\dotsc,t_n/v_n]\bigr]}$. Hence $\wp^\epsilon(t_1,\dotsc,t_n)$ belongs to $\LB\FLP\SELECT E\RB$; contradiction.
\end{proof}

The transformation of a formal logic program \FLP\ into a formal logic program of the form $\FLP\PLUS E$ will be of interest only in case $\Omega$ and $E$ are chosen in such a way that the condition in the definition that follows holds.

\begin{defn}\label{defn_47}
Let a formal logic program \FLP\ and a literal marker $\Omega$ for \FLP\ be given. We call an \emph{extensor for $(\FLP,\,\Omega)$} any set $E$ of literals over \VOCA\ such that $E\cup\LB\FLP\PLUS E\RB$ is consistent.
\end{defn}

Before we can end this section, we need one more technical notation. In relation to the well-founded semantics, it will be convenient to introduce an intermediate construction involving a family of extensors: a formal logic program \FLP\ will be extended to a formal logic program of the form $\FLP\PLUS E_0$, and then to a formal logic program of the form $(\FLP\PLUS E_0)+_{\Omega_1}E_1$, and then to a formal logic program of the form $(\FLP\PLUS E_0\cup E_1)+_{\Omega_2}E_2$, etc. Now $\Omega_1$, $\Omega_2$, etc., will not be arbitrary: they will all select occurrences of literals in $\FLP\PLUS E_0$, $\FLP\PLUS E_0\cup E_1$, etc., determined by $\Omega$, even though these occurrences of literals are taken from different formal logic programs as \FLP\ is being successively transformed. For instance, in Example~\ref{ex_42}, the occurrence of $\neg p_3(v_0)$ in $\varphi_{p_3}^+$ can be `tracked down' in \extends{E}{-}{\varphi_{p_3}^+}, though the (unique) occurrence of $\neg p_3(v_0)$ in $\varphi_{p_3}^+$ is of course different to the (unique) occurrence of $\neg p_3(v_0)$ in \extends{E}{-}{\varphi_{p_3}^+}. The following notation will allow us to formally express $\Omega_1$, $\Omega_2$, etc., from $\Omega$ and $E_0$, $E_1$, etc., and write $\Omega+E_0$ for $\Omega_1$, $\Omega+E_0\cup E_1$ for $\Omega_2$, etc. 

\begin{nota}\label{nota_48}
For all formulas $\varphi$, sets $O$ of occurrences of literals in $\varphi$
and nonsingleton members $o$ of $O$, let $\rho(O,\,o)$ be the set of occurrences $o'$ of literals in the formula in which $o\setminus\{\varphi\}$ is an occurrence of a literal, and such that $o'\cup\{\varphi\}\in O$.
Given a formula $\varphi$, a set $O$ of occurrences of literals in $\varphi$,
a set $E$ of literals and a member $o$ of $O$, set
\[
\extends{E}{O}{o}=
\begin{cases}
\{\extends{E}{O}{\varphi}\}\cup\extends{E}{\rho(O,\,o)}{o}\setminus\{\varphi\} &
\text{if $\varphi$ is not a literal},\\
\{\extends{E}{O}{\varphi},\,\varphi\} & \text{otherwise}.
\end{cases}
\]
\end{nota}

\begin{nota}\label{nota_49}
Let a formal logic program \FLP, a literal marker $\Omega$ for \FLP\, and a set $E$ of literals be given. Write $\Omega=(O_\wp^\epsilon)_{(\wp,\epsilon)\in\IND}$. We let $\Omega+E$ denote
\[
\bigl(\{\extends{E}{O_\wp^\epsilon}{o}\mid o\in O_\wp^\epsilon\}\bigr)_{(\wp,\epsilon)\in\IND}.
\]
\end{nota}

The property that follows justifies why the construction described before Notation~\ref{nota_48}  refers to a formal logic program of the form $(\FLP+_\Omega E_0\cup E_1)+_{\Omega_2}E_2$ rather than to a formal logic program of the form $\bigl((\FLP+_\Omega E_0)+_{\Omega_1}E_1\bigr)+_{\Omega_2}E_2$.

\begin{property}\label{property_50}
Let a formal logic program \FLP, a literal marker $\Omega$ for \FLP, and two sets $E$ and $F$ of literals be given. Then $\LB\FLP\PLUS E\cup F\RB=\LB(\FLP\PLUS E)+_{\Omega+E}F\RB$.
\end{property}

\subsection{Special extensors}\label{subsubsec_3_2}

The task of casting the well-founded, the stable model and the answer-set semantics into our framework boils down to defining appropriate literal markers and extensors. At a fundamental level, the question ``what are legitimate contextual assumptions?'' replaces the question ``how does negation behave?'' We now define the key properties that literal markers and extensors can enjoy and allow one to complete that task.

\begin{defn}\label{defn_51}
Let a formal logic program \FLP, a literal marker $\Omega$ for \FLP, and an extensor $E$ for $(\FLP,\,\Omega)$ be given.
\begin{itemize}
\item
We say that $E$ is \emph{imperative} iff for all closed literals $\varphi$,
\begin{center}
$\varphi$ is not an instance of a member of $E$ iff $\LB\FLP\PLUS E\RB\FORCES\SIM\varphi$.
\end{center}
\item
We say that $E$ is \emph{implicative} iff $E\subseteq\LB\FLP\PLUS E\RB$.
\item
We say that $E$ is \emph{supporting} iff for all $\psi\in E$, $\LB\FLP\RB\FORCES(\FLP\PLUS E)[\psi]$.
\item
Given an ordinal $\alpha$, we say that $E$ is \emph{$\alpha$-foundational}
iff there exists a sequence $(E_\beta)_{\beta<\alpha}$ of sets of literals such that $E=\bigcup_{\beta<\alpha}E_\beta$ and
for all ordinals $\beta<\alpha$,
$E_\beta$ is a supporting extensor for $(\FLP\PLUS\bigcup_{\gamma<\beta}E_\gamma,\,\Omega+\bigcup_{\gamma<\beta}E_\gamma)$.
\item
We say that $E$ is \emph{foundational} iff there exists a 
sequence $(E_\alpha)_{\alpha\in\ORD}$ of sets of literals such that $E=\bigcup_
{\alpha\in\ORD}E_\alpha$ and for all ordinals $\alpha$, $\bigcup_{\beta<\alpha}E_\beta$ is an $\alpha$-foundational extensor for $(\FLP,\,\Omega)$.
\end{itemize}
\end{defn}

Intuitively, an imperative extensor is a maximal set of hypotheses that will not be refuted, an implicative extensor is a set of hypotheses that will be confirmed, and a supporting extensor for \FLP\ is a set of hypotheses that will be confirmed thanks to themselves and to the literals generated by \FLP\ (but not to any nonhypothesis generated by a rule that fires only thanks to some hypothesis that activates its body). When the literal marker $\Omega$ marks all occurrences of literals that can unify with a hypothesis (so any hypothesis can be used in any context), supporting extensors have an alternative definition. This is what the next property expresses, with a corollary that will be used in relation to the well-founded semantics.

\begin{property}\label{property_52}
Let a formal logic program $\FLP=(\varphi_\wp^\epsilon)_{(\wp,\epsilon)\in\IND}$, a literal marker $\Omega=(O_\wp^\epsilon)_{(\wp,\epsilon)\in\IND}$ for \FLP, and an extensor $E$ for $(\FLP,\,\Omega)$ be such that for all $n\in\NSET$, $\wp\in\PRDVN{n}$, $\epsilon\in\{+,-\}$, literals $\psi$ and occurrences $o$ of $\psi$ in $\varphi_\wp^\epsilon$, if some closed instance of $\psi$ is an instance of a member of $E$ then $o\in O_\wp^\epsilon$. Then $E$ is supporting iff for all $\psi\in E$, $\LB\FLP\RB\cup E\FORCES\FLP[\psi]$.
\end{property}

\begin{cor}\label{cor_53}
Let a formal logic program \FLP\ be given. Let a supporting extensor $E$ for $(\FLP,\,\SPE{\cdot,-})$ consist of  negated atoms only. Then $E$ is supporting iff for all $\psi\in E$, $\LB\FLP\RB\cup E\FORCES\FLP[\psi]$.
\end{cor}

Recall that we have defined a set $X$ of literals to be saturated iff every closed atom is an instance of a member of at least one of the sets $X$ and $\SIM X$.

\begin{property}\label{property_54}
For all formal logic programs \FLP\ and literal markers $\Omega$ for \FLP, all imperative extensors for $(\FLP,\,\Omega)$ are saturated.
\end{property}

\begin{property}\label{property_55}
For all formal logic programs \FLP\ and literal markers $\Omega$ for \FLP, all supporting extensors for $(\FLP,\,\Omega)$ are implicative.
\end{property}

\begin{property}\label{property_56}
Let a formal logic program \FLP, a literal marker $\Omega$ for \FLP, and an extensor $E$ for $(\FLP,\,\Omega)$ be given.
\begin{itemize}
\item
For all ordinals $\alpha$, if $E$ is $\alpha$-foundational then $E$ is foundational.
\item
For all ordinals $\alpha$, if $E$ is $\alpha$-foundational then $E$ is $\beta$-foundational for all ordinals $\beta>\alpha$.
\item
If $E$ is foundational then there is $\alpha\in\ORD$ such that $E$ is $\alpha$-foundational.
\end{itemize}
\end{property}

It will be shown that the well-founded semantics is related to foundational extensors, and the answer-set semantics to imperative extensors. As for the stable model semantics, it will be shown to be related to both imperative and implicative extensors, by virtue of the following property.

\begin{property}\label{property_57}
For all formal logic programs \FLP, literal markers $\Omega$ for \FLP\ and complete sets $E$ of literals over \VOCA, $E$ is an implicative extensor for $(\FLP,\,\Omega)$ iff $E$ is an imperative extensor for $(\FLP,\,\Omega)$.
\end{property}

It is fair to say that to cast the answer-set, the stable model and the well-founded semantics into our framework, it would be sufficient to work under the assumption that $\VOCA=\VOCAB$: either these semantics are developed in a propositional setting, or they restrict the class of interpretations to Herbrand structures. There is no need to impose such restrictions, but a natural question is how much more general the notions become when the equality $\VOCA=\VOCAB$ is not imposed. In relation to the answer-set and the stable model semantics, the answer is: not much more. Indeed, the following proposition establishes that when \VOCA\ and \VOCAB\ are distinct, the notion of imperative extensor is often degenerate.

\begin{prop}\label{prop_58}
Suppose that $\VOCAB\setminus\VOCA$ contains a function symbol of arity 1 at least.
Let a formal logic program \FLP, a literal marker $\Omega$ for \FLP, and an imperative extensor $E$ for $(\FLP,\,\Omega)$ be given. Then for all $n\in\NSET$ and $\wp\in\PRDVN{n}$, the set of members of $\LB\FLP\PLUS E\RB$ of the form $\wp^\epsilon(t_1,\dotsc,t_n)$ is either empty or equal to the set of all atoms over \VOCA\ of the form $\wp(t_1,\dotsc,t_n)$ or equal to the set of all negated atoms over \VOCA\ of the form $\neg\wp(t_1,\dotsc,t_n)$.
\end{prop}

\begin{proof}
There is nothing to prove if \VOCAB\ contains no constant, so suppose otherwise.
Let $n\in\NSET$ and $\wp\in\PRDVN{n}$ be given.
\begin{itemize}
\item
Let $X$ be the set of $n$-tuples of terms over \VOCA, say $(t_1,\dotsc,t_n)$, such that for all closed terms $t'_1$, \dots, $t'_n$, if $(t'_1,\dotsc,t'_n)$ is an instance of $(t_1,\dotsc,t_n)$ then both $\wp(t'_1,\dotsc,t'_n)$ and $\neg\wp(t'_1,\dotsc,t'_n)$ are instances of members of $E$. 
\item
For all $\epsilon\in\{+,-\}$, let $X^\epsilon$ be the set of $n$-tuples of terms over \VOCA, say $(t_1,\dotsc,t_n)$, such that $\wp^\epsilon(t_1,\dotsc,t_n)\in\LB\FLP\PLUS E\RB$.
\end{itemize}
Using the fact that $E$ is an imperative extensor for $(\FLP,\,\Omega)$, it is easy to verify that $X$, $X^+$ and $X^-$ are disjoint and that for all closed terms $t_1$, \dots, $t_n$, $(t_1,\dotsc,t_n)$ is an instance of some member of $X\cup X^+\cup X^-$.
Let a nonnullary function symbol $f$ in $\VOCAB\setminus\VOCA$ be given. Then there exists an $n$-tuple $(\iota_1,\dotsc,\iota_n)$ of distinct closed terms that all start with $f$. Obviously, for all terms $t_1$, \dots, $t_n$ over \VOCA, if $(\iota_1,\dotsc,\iota_n)$ is an instance of $(t_1,\dotsc,t_n)$ then $t_1$, \dots, $t_n$ are distinct variables.
So
either all $n$-tuples of closed terms are instances of some member of $X$, in which case $\LB\FLP\PLUS E\RB$ contains no literal over \VOCA\ of the form $\wp(t_1,\dotsc,t_n)$ or $\neg\wp(t_1,\dotsc,t_n)$,
or all $n$-tuples of closed terms are instances of some member of $X^+$, in which case $\LB\FLP\PLUS E\RB$ contains all literals over \VOCA\ of the form $\wp(t_1,\dotsc,t_n)$,
or all $n$-tuples of closed terms are instances of some member of $X^-$, in which case $\LB\FLP\PLUS E\RB$ contains all literals over \VOCA\ of the form $\neg\wp(t_1,\dotsc,t_n)$,
completing the proof of the proposition.
\end{proof}

The following example shows that if $\VOCAB\setminus\VOCA$ does not contain a function symbol of arity 1 at least, then the notion of imperative extensor can be nondegenerate.

\begin{ex}\label{ex_59}
Suppose that \VOCAB\ consists of \NUM{0}, $s$ and a binary predicate symbol $p$, and assume that $\VOCA=\{s,\,p\}$. Let \FLP\ be the formal logic program determined by $\varphi_p^+\equiv v_1\doteq v_2$ and $\varphi_p^-\equiv\bigvee\varnothing$. Let $E$ be the set of literals defined as
\[
\bigl\{p(v_1,v_2)\bigr\}\cup\bigl\{\neg p(s^n(v_0),v_0),\,\neg p(v_0,s^n(v_0))\bigm| n\in\NSET\setminus\{0\}\bigr\}.
\]
Set $\Omega=(\varnothing,\,\varnothing)$.
Then $E$ is an imperative extensor for $(\FLP,\,\Omega)$ and $\LB\FLP\PLUS E\RB$, which is obviously equal to $\LB\FLP\RB$, is $\bigl\{p(s^n(v_i),s^n(v_i))\mid n\in\NSET,\,i\in\NSET\bigr\}$.
\end{ex}

To summarise the previous considerations, we have not assumed in Definition~\ref{defn_51} that \VOCA\ and \VOCAB\ are equal simply because none of the results we want to establish needs that assumption to be made. But the notion of imperative extensor (which is the key notion in relation to the stable model and the answer-set semantics) is defined in such a way that it is only interesting when $\VOCA=\VOCAB$ or when \VOCA\ and \VOCAB\ take very specific values.

\subsection{A few technical results}

The results that follow will be used in the sequel.

\begin{lem}\label{lem_60}
Let a formal logic program \FLP\ and a literal marker $\Omega$ for \FLP\ be given.
For all sets $E$ and $F$ of literals, if $E\subseteq F$
then $\LB\FLP\PLUS E\RB\subseteq\LB\FLP\PLUS F\RB$.
\end{lem}

\begin{proof}
Let $E$ and $F$ be two sets of literals with $E\subseteq F$.
It is immediately verified by induction that for all ordinals $\alpha$, $\LB\FLP\PLUS E\RB_\alpha\subseteq\LB\FLP\PLUS F\RB_\alpha$. We conclude with Property~\ref{property_26}.
\end{proof}

\begin{lem}\label{lem_61}
Let a formal logic program \FLP\ and a literal marker $\Omega$ for \FLP\ be given.
For all sets $E$ and $F$ of literals, $\BLB\FLP\PLUS\LB\FLP\PLUS E\RB\cup F\BRB\subseteq\LB\FLP\PLUS E\cup F\RB$.
\end{lem}

\begin{proof}
Let $E$ and $F$ be two sets of literals. Let ordinal $\lambda$ be such that $\LB\FLP\PLUS E\RB_\lambda$ is equal to $\LB\FLP\PLUS E\RB_{\lambda+1}$. 
It is easy to verify by induction that for all ordinals $\alpha$, $\LB\FLP\PLUS\LB\FLP\PLUS E\RB\cup F\RB_\alpha\subseteq\LB\FLP\PLUS E\cup F\RB_{\lambda+\alpha}$. We conclude with Property~\ref{property_26}.
\end{proof}

\begin{lem}\label{lem_62}
For all formal logic programs \FLP, literal markers $\Omega$ for \FLP\ and implicative extensors $E$ for $(\FLP,\,\Omega)$, $\BLB\FLP\PLUS\LB\FLP\PLUS E\RB\BRB=\LB\FLP\PLUS E\RB$.
\end{lem}

\begin{proof}
The lemma follows immediately from Lemmas~\ref{lem_60} and~\ref{lem_61}.
\end{proof}

\begin{cor}\label{cor_63}
For all formal logic programs \FLP, literal markers $\Omega$ for \FLP\ and implicative extensors $E$ for $(\FLP,\,\Omega)$, $\LB\FLP\PLUS E\RB$ is an implicative extensor for $(\FLP,\,\Omega)$.
\end{cor}

\begin{prop}\label{prop_64}
Let a formal logic program \FLP\ be locally consistent. Let a literal marker $\Omega$ for \FLP\ be given. Let a set $X$ of implicative extensors for $(\FLP,\,\Omega)$ be such that $\bigcup X$ is consistent. Then $\bigcup X$ is an extensor for $(\FLP,\,\Omega)$.
\end{prop}

\begin{proof}
Write $\FLP=(\varphi_\wp^\epsilon)_{(\wp,\epsilon)\in\IND}$.
Set $E=\bigcup X$.
We show by induction that for all ordinals $\alpha$,
$E\cup\LB\FLP\PLUS E\RB_\alpha$ is consistent. Let an ordinal $\alpha$ be given and assume that for all $\beta<\alpha$, $E\cup\LB\FLP\PLUS E\RB_\beta$ is consistent.
Since \FLP\ is locally consistent and $E$ is consistent (used in the case where $\alpha=0$), there exists no $n\in\NSET$, $\wp\in\PRDVN{n}$ and closed terms $t_1$, \dots, $t_n$ such that $E\cup\bigcup_{\beta<\alpha}\LB\FLP\PLUS E\RB_\beta$ forces $\varphi_\wp^+[t_1/v_1,\dotsc,t_n/v_n]$ and $\varphi_\wp^-[t_1/v_1,\dotsc,t_n/v_n]$. Hence $E\cup\LB\FLP\PLUS E\RB_\alpha$ cannot be inconsistent unless the set of closed instances of members of $\LB\FLP\PLUS E\RB_\alpha$ intersects the set of closed instances of members of $\SIM E$. Assume that the set of closed instances of members of $\LB\FLP\PLUS E\RB_\alpha$ indeed intersects the set of closed instances of members of $\SIM E$. Since all members of $X$ are implicative, any closed instance of any member of $E$ is an instance of some member of $\bigcup_{F\in X}\LB\FLP\PLUS F\RB$. Let ordinal $\lambda$ be least such that there exists a closed literal $\varphi$ with $\bigcup_{F\in X}\LB\FLP\PLUS F\RB_\lambda\FORCES\varphi$ and $\LB\FLP\PLUS E\RB_\alpha\FORCES\SIM\varphi$.
Let $F\in X$ and a closed literal $\varphi$ be such that $\LB\FLP\PLUS F\RB_\lambda\FORCES\varphi$ and $\LB\FLP\PLUS E\RB_\alpha\FORCES\SIM\varphi$.
Set
\[
Y=\bigcup_{\beta<\lambda}\LB\FLP\PLUS F\RB_\beta\cup\LB\FLP\PLUS E\RB_\alpha.
\]
We derive from the choice of $\lambda$ that
$Y$ is consistent. Let $n\in\NSET$, $\wp\in\PRDVN{n}$ and terms $t_1$, \dots, $t_n$
be such that $\varphi$ is $\wp(t_1,\dotsc,t_n)$ or $\neg\wp(t_1,\dotsc,t_n)$.
By the choice of $\varphi$, $Y$ forces both $\varphi_\wp^+[t_1/v_1,\dotsc,t_n/v_n]$ and
$\varphi_\wp^-[t_1/v_1,\dotsc,t_n/v_n]$, which is impossible since \FLP\ is locally consistent. We conclude that $E\cup\LB\FLP\PLUS E\RB_\alpha$ is consistent.
\end{proof}

As an immediate consequence of Property~\ref{property_55} and Proposition~\ref{prop_64}:

\begin{cor}\label{cor_65}
Let a formal logic program \FLP\ be locally consistent.
Let a literal marker $\Omega$ for \FLP\ be given.
Let $X$ be a set of supporting extensors for $(\FLP,\,\Omega)$ such that
$\bigcup X$ is consistent. Then $\bigcup X$ is a supporting extensor for $(\FLP,\,\Omega)$.
\end{cor}

To end this section, let us give a simple application of some of the previous observations.
Complete sets of literals can obviously be identified with standard structures, hence it is 
natural to ask whether a complete set of the form $\LB\FLP\PLUS E\RB$ is a 
model of the classical logical form of \FLP. It is easy
to answer that question positively for implicative extensors.

\begin{prop}\label{prop_66}
Let a formal logic program \FLP, a literal marker $\Omega$ for \FLP, and an implicative extensor $E$ for $(\FLP,\,\Omega)$ be such that $\LB\FLP\PLUS E\RB$ is complete.
Then the set of closed instances of atoms in $\LB\FLP\PLUS E\RB$ is a model of $\CLF(\FLP)$.
\end{prop}

\begin{proof}
Obviously, for all formulas $\varphi$ and sets $O$ of occurrences of literals in $\varphi$, $\varphi$ logically implies \extends{\FLP\PLUS\LB\FLP\PLUS E\RB}{O}{\varphi} in \WORLDS.
It follows that $\CLF(\FLP\PLUS\LB\FLP\PLUS E\RB)$ logically implies $\CLF(\FLP)$ in \WORLDS.
By Lemma~\ref{lem_62}, $\LB\FLP\PLUS E\RB=\BLB\FLP\PLUS\LB\FLP\PLUS E\RB\BRB$, and we derive from Corollary~\ref{cor_33} that $\LB\FLP\PLUS E\RB$ logically implies $\CLF(\FLP\PLUS\LB\FLP\PLUS E\RB)$ in \WORLDS.
We conclude that $\LB\FLP\PLUS E\RB\WMODELS\CLF(\FLP)$.
\end{proof}

\subsection{Relationship to the answer-set semantics}\label{subsubsec_4_2_1}

In this section we consider the enrichment of \INFLVOCAB\ with a second negation operator, written \NOT,
that can be applied to any literal, and to literals only.
We do not develop the formalism beyond this minimalist syntactic
consideration as we use \NOT\ to remind the reader of the usual definition of answer-sets, but we will not use it in an alternative definition of answer-sets that will immediately be seen to be equivalent to the usual definition. For this purpose, let us introduce some preliminary notation. Let a formula $\varphi$ and a set $O$ of occurrences of literals in $\varphi$ be given. We define a member $\varphi[O]$ of the enrichment of \INFLVOCAB\ with \NOT, thanks to the inductive construction that follows.
\begin{itemize}
\item
Suppose that $\varphi$ is an identity, a distinction, or a literal.
\begin{itemize}
\item
If $O=\varnothing$ then $\varphi[O]$ is $\varphi$.
\item
If $O=\{\varphi\}$ and $\varphi$ is an atom then $\varphi[O]$ is $\NOT\neg\varphi$.
\item
If $O=\{\varphi\}$ and $\varphi$ is of the form $\neg\psi$ then $\varphi[O]$ is $\NOT\psi$.
\end{itemize}
\item
Suppose that $\varphi$ is of the form $\bigvee X$ or $\bigwedge X$. For all $\psi\in X$, let
$O_\psi$ be the (unique) set of occurrences of literals in $\psi$, say $o$, with $o\cup\{\varphi\}\in O$.
\begin{itemize}
\item
If $\varphi$ is the formula $\bigvee X$ then $\varphi[O]$ is $\bigvee\{\psi[O_\psi]\mid\psi\in X\}$.
\item
If $\varphi$ is the formula $\bigwedge X$ then $\varphi[O]$ is $\bigwedge\{\psi[O_\psi]\mid\psi\in X\}$.
\end{itemize}
\item
Suppose that $\varphi$ is of the form $\exists x\,\psi$ or $\forall x\,\psi$.
Let $O_\psi$ be the (unique) set of occurrences of literals in $\psi$, say $o$, with $o\cup\{\varphi\}\in O$.
\begin{itemize}
\item
If $\varphi$ is the formula $\exists x\,\psi$ then $\varphi[O]$ is $\exists x\,\psi[O_\psi]$.
\item
If $\varphi$ is the formula $\forall x\,\psi$ then $\varphi[O]$ is $\forall x\,\psi[O_\psi]$.
\end{itemize}
\end{itemize}
Now let a formal logic program \FLP\ and a literal marker $\Omega$ for \FLP\ be given.
Write $\FLP=(\varphi_\wp^\epsilon)_{(\wp,\epsilon)\in\IND}$ and $\Omega=(O_\wp^\epsilon)_{(\wp,\epsilon)\in\IND}$, and set $\FLP[\Omega]=\bigl(\varphi_\wp^\epsilon[O_\wp^\epsilon]\bigr)_{(\wp,\epsilon)\in\IND}$. Then $\FLP[\Omega]$ is what is known in the literature as an extended logic program, a logic program with two kinds of negation, $\neg$ and \NOT. 
Conversely, let an extended logic program $G$ that, without loss of generality, is written in such 
a way that for every $n\in\NSET$ and $\wp\in\PRDVN{n}$, $G$ has one rule whose head is 
$\wp(v_1,\dotsc,v_n)$, one rule whose head is $\neg\wp(v_1,\dotsc,v_n)$, and no other rule 
whose head is of the form $\wp(t_1,\dotsc,t_n)$ or $\neg\wp(t_1,\dotsc,t_n)$. Then there exists 
a unique formal logic program \FLP\ and a unique literal marker $\Omega$ for \FLP\ 
with $G=\FLP[\Omega]$.

For instance, assume that \VOCAB\ consists of 4 nullary predicate symbols $p_1$, $p_2$, 
$p_3$ and $p_4$. Suppose that \FLP\ is given by the following formulas.
\begin{align*}
&\varphi_{p_1}^+\equiv p_2\wedge p_3& &\varphi_{p_2}^+\equiv p_4& &\varphi_{p_3}^+\equiv p_3& &
\varphi_{p_4}^+\equiv \neg p_3\\
&\varphi_{p_1}^-\equiv p_2\vee\neg p_4& &\varphi_{p_2}^-\equiv \neg p_3& &\varphi_{p_3}^-\equiv \neg p_3
\wedge p_2& &\varphi_{p_4}^-\equiv \textstyle\bigvee\varnothing
\end{align*}
Suppose that $\Omega$ is given by the following sets.
\begin{align*}
&O_{p_1}^+\equiv \varnothing& &O_{p_1}^-\equiv \bigl\{\{p_2\vee\neg p_4,\,p_2\},\,\{p_2\vee\neg p_4,\,\neg p_4\}\bigr\}\\
&O_{p_2}^+\equiv \varnothing& &O_{p_2}^-\equiv \varnothing\\
&O_{p_3}^+\equiv \varnothing& &O_{p_3}^-\equiv \bigl\{\{\neg p_3\wedge p_2,\,p_2\}\bigr\}\\
&O_{p_4}^+\equiv \bigl\{\{\neg p_3\}\bigr\}& &O_{p_4}^-\equiv \varnothing
\end{align*}
So $(\FLP,\,\Omega)$ can be represented as
\begin{align*}
p_1&\leftarrow p_2\wedge p_3& p_2&\leftarrow p_4& p_3&\leftarrow p_3& p_4&\leftarrow\underset{\checkmark}{\neg p_3}\\
\neg p_1&\leftarrow\underset{\checkmark}{p_2}\vee\underset{\checkmark}{\neg p_4}& \neg p_2&\leftarrow \neg p_3& \neg p_3&\leftarrow\neg p_3\wedge\underset{\checkmark}{p_2}
\end{align*}
and $\FLP[\Omega]$ is the extended logic program
\begin{align*}
p_1&\leftarrow p_2\wedge p_3& p_2&\leftarrow p_4& p_3&\leftarrow p_3& p_4&\leftarrow\NOT p_3\\
\neg p_1&\leftarrow\NOT\neg p_2\vee\NOT p_4& \neg p_2&\leftarrow \neg p_3& \neg p_3&\leftarrow\neg p_3\wedge\NOT\neg p_2
\end{align*}
Moreover, it is easy to see that the extensors for $(\FLP,\,\Omega)$ are:
\begin{itemize}
\item
all subsets $E$ of $\{p_1,\,\neg p_1,\,\neg p_2,\,p_3,\,p_4\}$, in which case $
\LB\FLP\PLUS E\RB=\varnothing$;
\item
all subsets $E$ of $\{p_1,\,\neg p_1,\,p_2,\,\neg p_2,\,p_3,\,p_4,\,\neg p_4\}$ which contain at least one of $p_2$ and $\neg p_4$, in which case $\LB\FLP\PLUS E\RB=\{\neg p_1\}$;
\item
all subsets $E$ of $\{p_1,\,\neg p_1,\,p_2,\,\neg p_2,\,p_3,\,\neg p_3,\,p_4,\,\neg p_4\}$ which $\neg p_3$ belongs to, in which case $
\LB\FLP\PLUS E\RB=\{\neg p_1,\,p_2,\,p_4\}$.
\end{itemize}
Out of these, only $\{\neg p_1,\,p_2,\,p_3,\,\neg p_3,\,p_4\}$ is imperative. 
Moreover, there is a unique answer-set for $\FLP[\Omega]$, namely $\{\neg p_1,\,p_2,\,p_4\}$.

Having realised that the class of extended logic programs is in one-to-one correspondence with the class of pairs $(\FLP,\,\Omega)$ where \FLP\ is a formal logic program and $\Omega$ a literal marker for \FLP\ (the correspondence in question putting a pair of the form $(\FLP,\,\Omega)$ in relation to $\FLP[\Omega]$), it is easy to see that if one assumes that \VOCA\ is equal to \VOCAB, then Definition~\ref{defn_67} amounts to the notion of an answer-set---recall the discussion at the end of Section~\ref{subsubsec_3_2} about not assuming that \VOCA\ and \VOCAB\ are equal.

\begin{defn}\label{defn_67}
Let a formal logic program \FLP\ and a literal marker $\Omega$ for \FLP\ be given. An \emph{answer-set for $(\,\FLP,\,\Omega\,)$} is a partial interpretation $M$ for which there exists a (necessarily saturated) set $E$ of literals over \VOCA\ with the following property.
\begin{itemize}
\item
For all closed literals $\varphi$, $\varphi\in M$ iff $\SIM\varphi$ is not an instance of a member of $E$.
\item
$M$ is the set of closed instances of members of $\LB\FLP\SELECT E\RB$.
\end{itemize}
\end{defn}

The next proposition shows that the concept of imperative extensor fully characterises the notion of answer-set.

\begin{prop}\label{prop_68}
Let a formal logic program \FLP, a literal marker $\Omega$ for \FLP, and a set $E$ of literals over \VOCA\ be given.
Let $F$ be the set of closed instances of members of $E$, and let $M$ be the set of all closed literals $\varphi$ with $\SIM\varphi\notin F$. Then $E$ is an imperative extensor for $(\,\FLP,\,\Omega\,)$ iff $M$ is an answer-set for $(\,\FLP,\,\Omega\,)$.
\end{prop}

\begin{proof}
Assume that $E$ is an imperative extensor for $(\,\FLP,\,\Omega\,)$. By Definition~\ref{defn_51}, the set of closed instances of members of $\LB\FLP\PLUS E\RB$ is consistent, is precisely equal to $M$, and is included in $F$, which implies by Lemma~\ref{lem_46} that $\LB\FLP\PLUS E\RB=\LB\FLP\SELECT E\RB$. We conclude that $M$ is an answer-set for $(\,\FLP,\,\Omega\,)$.

Conversely, assume that $M$ is an answer-set for $(\FLP,\,\Omega)$. Hence $M$ is consistent, and so $M\subseteq F$.
By Definition~\ref{defn_67} and Property~\ref{property_45}, $M$ consists of the closed instances of the members of $\LB\FLP\SELECT E\RB$, and so by Lemma~\ref{lem_46}, consists of the closed instances of the members of $\LB\FLP\PLUS E\RB$. Hence $E$ is an imperative extensor for $(\,\FLP,\,\Omega\,)$.
\end{proof}

In the answer-set semantics, $\NOT\varphi$ intuitively means that $\varphi$ is not provable, that is, not derived. 
The way to go from the usual presentation of the answer-set semantics to our setting is to let  hypotheses of the form $\SIM\varphi$  take effect in contexts where the answer-set framework has statements of the form ``$\varphi$ is not provable.''
The fact that $\varphi$ is either provable or not is then mapped to the constraint, captured by the notion of imperative extensor, that either $\varphi$ should be derived or $\SIM\varphi$ should be assumed.

\subsection{Relationship to the stable model semantics}

The stable model semantics takes the sets of positive rules as the object of study; but as mentioned repeatedly, the class of these sets is in one-to-one correspondence with the class of
symmetric formal logic programs, hence it is legitimate to study the stable model semantics on the basis of the latter. If one assumes that \VOCA\ is equal to \VOCAB, then Definition~\ref{defn_69} captures the notion of stable model---again, recall the discussion at the end of Section~\ref{subsubsec_3_2} about not assuming that \VOCA\ and \VOCAB\ are equal. Note how Notation~\ref{nota_40} is being used in Definition~\ref{defn_69} to basically describe the Lloyd-Topor transformation.

\begin{defn}\label{defn_69}
Let a formal logic program $\FLP=(\varphi_\wp^\epsilon)_{(\wp,\epsilon)\in\IND}$ be given. A partial interpretation $M$ is said to be \emph{stable for \FLP} iff there exists a complete set $E$ of literals over \VOCA\ such that $M$ is the set of closed instances of members of $E$ and for all closed atoms $\varphi$, $\varphi\in M$ iff
\[
\bigl\{\selects{E}{-}{\varphi_\wp^+}\rightarrow\wp(v_1,\dotsc,v_n)\bigm| n\in\NSET,\,\wp\in\PRDVN{n}\bigr\}\WMODELS\varphi.
\]
\end{defn}

Note that the condition on $E$ only depends on the positive rules of \FLP. In Definition~\ref{defn_69}, \FLP\ is not assumed to be symmetric; but it is essential to assume that \FLP\ is symmetric to obtain the result stated in the proposition that follows. Together with Property~\ref{property_57}, this proposition shows that both concepts of imperative and implicative extensors relative to the literal markers that collect all occurrences of all negated atoms fully characterise the notion of stable model.

\begin{prop}\label{prop_70}
For all symmetric formal logic programs \FLP\ and complete sets $E$ of literals over \VOCA, the set of closed instances of members of $E$ is stable for \FLP\ iff $E$ is an implicative extensor for $(\FLP,\,\SPE{-,-})$.
\end{prop}

\begin{proof}
Let a symmetric formal logic program $\FLP=(\varphi_\wp^\epsilon)_{(\wp,\epsilon)\in\IND}$ and a complete set $E$ of literals over \VOCA\ be given. Let $E^+$ be the set of atoms in $E$, and let $E^-$ be the set of negated atoms in $E$.

Suppose that the set of closed instances of members of $E$ is stable for \FLP.
Then $\CLF(\FLP\SELECTNEG E)$ logically implies $E^+$ in \WORLDS.
Since negation does not occur in the left hand side of any implication in $\CLF(\FLP\SELECTNEG E)$, it follows that $E^+$ is a subset of $\LB\FLP\SELECTNEG E\RB$.
Let $n\in\NSET$, $\wp\in\PRDVN{n}$, and closed terms
$t_1$, \dots, $t_n$ be given.
Since $E$ is complete and \FLP\ is symmetric, $E$ forces one and only one of $\varphi_\wp^+[t_1/v_1,\dotsc,t_n/v_n]$ and
$\varphi_\wp^-[t_1/v_1,\dotsc,t_n/v_n]$.
Suppose that $\neg\wp(t_1,\dotsc,t_n)$ is an instance of a member of $E$. If $E\FORCES\varphi_\wp^+[t_1/v_1,\dotsc,t_n/v_n]$ then $E^+\FORCES\selects{E}{-}{\varphi_\wp^+}[t_1/v_1,\dotsc,t_n/v_n]$, hence there exists terms $t'_1$, \dots, $t'_n$ over \VOCA\ such that $(t_1,\dotsc,t_n)$ is an instance of $(t'_1,\dotsc,t'_n)$ and $\wp(t'_1,\dotsc,t'_n)\in E$, contradicting the assumption that $E$ is consistent. We infer that $E^+\FORCES\selects{E}{-}{\varphi_\wp^-}[t_1/v_1,\dotsc,t_n/v_n]$, hence there exists terms $t'_1$, \dots, $t'_n$ over \VOCA\ such that $(t_1,\dotsc,t_n)$ is an instance of $(t'_1,\dotsc,t'_n)$ and $\neg\wp(t'_1,\ldots,t'_n)$ belongs to $\LB\FLP\SELECTNEG E\RB$.
Suppose that $\wp(t_1,\dotsc,t_n)$ is an instance of a member of $E$. Then $E\FORCES\varphi_\wp^+[t_1/v_1,\dotsc,t_n/v_n]$, hence
$E\NFORCES\varphi_\wp^-[t_1/v_1,\dotsc,t_n/v_n]$, hence $E$ does not force
$\selects{E}{-}{\varphi_\wp^-}[t'_1/v_1,\dotsc,t'_n/v_n]$ for any terms $t'_1$, \dots, $t'_n$ over \VOCA\ such that $(t_1,\dotsc,t_n)$ is an instance of $(t'_1,\dotsc,t'_n)$. It is then easy to conclude that for all closed literals $\varphi$, $\LB\FLP\SELECTNEG E\RB\FORCES\varphi$ iff $\varphi$ is an instance of a member of $E$. Together with Lemma~\ref{lem_46},
this completes the verification that $E$ is an implicative extensor for $(\FLP,\,\SPE{-,-})$.

Conversely, assume that $E$ is an implicative extensor for $(\FLP,\,\SPE{-,-})$.
Since $E$ is complete, Lemma~\ref{lem_46} again implies that $\LB\FLP\PLUSNEG E\RB=\LB\FLP\SELECTNEG E\RB$.
Set
\[
X=\bigl\{\selects{E}{-}{\varphi_\wp^+}\rightarrow\wp(v_1,\dotsc,v_n)\bigm| \\
 n\in\NSET,\,\wp\in\PRDVN{n}\bigr\}.
\]
Clearly, $\CLF(\FLP\SELECTNEG E)$, being logically equivalent in \WORLDS\ to the complete set $E$, is also logically equivalent to $E^-\cup X$ in \WORLDS. Hence $E^-\cup X\WMODELS E^+$. Since negation does not occur in any implication
in $X$, this implies that $X\WMODELS E^+$, which completes
the verification that the set of closed instances of members of $E$ is stable for \FLP.
\end{proof}

\subsection{Supporting and foundational extensors}

The well-founded semantics is related to the notion of foundational extensor, and we will need to establish some of the properties that the latter enjoys in order to establish the relationship. The notion of supporting extensor has mainly been introduced as a useful building block in the definition of foundational extensors, but it is interesting in its own right. By Property~\ref{property_55}, supporting  extensors are implicative extensors, which means that they consist of hypotheses that are guaranteed to be confirmed. But more is true. Intuitively, given a formal logic program \FLP\ and a literal marker $\Omega$ for \FLP,
a supporting extensor for $(\FLP,\,\Omega)$ is sufficiently rich in literals
to `generate itself' using \FLP\ and $\Omega$, and not contradict any literal generated by $\FLP\PLUS E$. So for all members $\varphi$ of
a supporting extensor $E$ for $(\FLP,\,\Omega)$, there exists a `constructive
proof' of $\varphi$, from the rules formalised as \FLP, such that the only literals that occur in the proof either are in $\LB\FLP\RB$ or are members of $E$ that occur in contexts where $\Omega$ accepts that they be assumed.
The next example will help grasp the idea in the simple case where $\Omega$ accepts that any literal be assumed in any context, and where no member of $\LB\FLP\RB$ is actually needed in the `constructive proofs'.

\begin{ex}\label{ex_71}
If \FLP\ is the formal logic program of Example~\ref{ex_17} then the supporting 
extensors for $(\FLP,\,\SPE{\pm,\pm})$ which are disjoint from $\LB\FLP\RB$ are the consistent unions of
\begin{itemize}
\item
$\{p_4(\NUM{2n}) \mid n\geq m\}$ where $m$ is an arbitrary member of \NSET,
\item
$\{\neg p_4(\NUM{2n}) \mid n\geq m\}$ where $m$ is an arbitrary member of \NSET,
\item
$\{p_4(\NUM{2n+1}) \mid n\geq m\}$ where $m$ is an arbitrary member of \NSET,
\item
$\{\neg p_4(\NUM{2n+1}) \mid n\geq m\}$ where $m$ is an arbitrary member of \NSET,
\item
$\{q_2,\,q_3\}$,
\item
$\{\neg q_2,\,\neg q_3\}$, and
\item
$\{q_4\}$.
\end{itemize}
\end{ex}

Casting the well-founded semantics into our framework requires to focus on symmetric formal logic programs only. But the notion of $\subseteq$-maximal foundational extensor, which will be seen to formalise the key principle behind the well-founded semantics, can be applied to arbitrary formal logic programs, hence to \FLP\ of Example~\ref{ex_17}. For this particular formal logic program, the notion of $\subseteq$-maximal foundational extensor reduces to that of $\subseteq$-maximal supporting extensor; this is because in this particular case, the process of transfinitely transforming \FLP\ with a $\subseteq$-maximal supporting extensor converges after its first application. Also, because of its full bias towards negated atoms, the well-founded semantics elects the literal marker that marks all occurrences of all negated atoms in the bodies of all rules or, alternatively, all negated atoms in the bodies of all negative rules (both choices are equivalent). It will be seen that the well-founded semantics of \FLP\ is captured by the $\subseteq$-maximal set of negated atoms over \VOCA\ that is a foundational extensor for $(\FLP,\,\SPE{-,-})$ or $(\FLP,\,\SPE{\cdot,-})$; with respect to the previous example, that set is
$\{\neg p_4(\NUM{n}) \mid n\in\NSET\}\cup\{\neg q_2,\,\neg q_3\}$.

The `dual' of that semantics would be fully biased towards nonnegated atoms, and would be captured by the $\subseteq$-maximal set of atoms over \VOCA\ that is a foundational extensor for $(\FLP,\,\SPE{+,+})$ or $(\FLP,\,\SPE{+,\cdot})$ (where \SPE{+,+} and \SPE{+,\cdot} would denote the literal marker for \FLP\ that marks all occurrences of all nonnegated atoms in the bodies of all rules or all positive rules, respectively); in the context of the previous example, that set is
$\{p_4(\NUM{n}) \mid n\in\NSET\}\cup\{q_2,\,q_3,\,q_4\}$.

A `balanced' semantics in the family of the semantics determined by maximal foundational extensors could elect a $\subseteq$-maximal foundational extensor for $(\FLP,\,\SPE{\pm,\pm})$ that contains
$\{p_4(\NUM{2n}),\,\neg p_4(\NUM{2n+1}) \mid n\in\NSET\}$,
and allows one to transform \FLP\ into a formal logic program that provides a fourth way of 
generating the set of even numbers and its complement, using the predicate symbol
$p_4$---besides the three options already available with $p_1$, $p_2$ and $p_3$. 

Subsuming the notion of foundational extensor given in Definition~\ref{defn_51} is the notion of foundational chain, that we make explicit in order to easily investigate
the properties of the foundational extensors.
Given a formal logic program \FLP\ and a literal marker $\Omega$ for $(\FLP,\,\Omega)$, a 
foundational chain for $(\FLP,\,\Omega)$ can be described as follows.
\begin{itemize}
\item
Start with a supporting extensor $E_0$ for $(\FLP,\,\Omega)$.
\item
Propose a supporting extensor $E_1$ for $(\FLP\PLUS E_0,\,\Omega+E_0)$.
\item
Propose a supporting extensor $E_2$ for $(\FLP\PLUS E_0\cup E_1,\,\Omega+E_0\cup E_1)$.
\item
Etc.
\end{itemize}

Formally, this translates into the following definition.

\begin{defn}\label{defn_72}
Let a formal logic program \FLP\ and a literal marker $\Omega$ for \FLP\ be given.

Given an ordinal $\alpha$, an \emph{$\alpha$-foundational chain
for\/ $(\FLP,\,\Omega)$} is a sequence $(E_\beta)_{\beta<\alpha}$ of sets of literals
over \VOCA\ such that for all ordinals $\beta<\alpha$, $E_\beta$
is a supporting extensor for $(\FLP\PLUS\bigcup_{\gamma<\beta}E_\gamma,\,\Omega+\bigcup_{\gamma<\beta}E_\gamma)$.

A \emph{foundational chain for\/ $(\FLP,\,\Omega)$} is a sequence
$(E_\alpha)_{\alpha\in\ORD}$ of sets of 
literals over \VOCA\ such that for all ordinals $\alpha$, $(E_\beta)_{\beta<\alpha}$ is
an $\alpha$-foundational chain for $(\FLP,\,\Omega)$.
\end{defn}

Let a formal logic program \FLP\ and a literal marker $\Omega$ for \FLP\ be given.
By Definitions~\ref{defn_51} and~\ref{defn_72},
\begin{itemize}
\item
for all $\alpha\in\ORD$ and
$\alpha$-foundational chains $(E_\beta)_{\beta<\alpha}$ for $(\FLP,\,\Omega)$,
$\bigcup_{\beta<\alpha}E_\beta$ is an $\alpha$-foundational
extensor for $(\FLP,\,\Omega)$ and
for all foundational chains $(E_\alpha)_{\alpha\in\ORD}$ for $(\FLP,\,\Omega)$,
$\bigcup_{\alpha\in\ORD}E_\alpha$ is a foundational extensor
for $(\FLP,\,\Omega)$;
\item
for all ordinals $\alpha$ and for all $\alpha$-foundational extensors $E$
for $(\FLP,\,\Omega)$, there exists an $\alpha$-foundational chain
$(E_\beta)_{\beta<\alpha}$ for $(\FLP,\,\Omega)$ such that
$E=\bigcup_{\beta<\alpha}E_\beta$ and
for all foundational extensors $E$ for $(\FLP,\,\Omega)$, there exists a foundational
chain $(E_\alpha)_{\alpha\in\ORD}$ for $(\FLP,\,\Omega)$ such that
$E=\bigcup_{\alpha\in\ORD}E_\alpha$.
\end{itemize}

The proposition that follows generalises Property~\ref{property_55}.

\begin{prop}\label{prop_73}
For all formal logic programs \FLP\ and literal markers $\Omega$ for \FLP,
all foundational extensors for $(\FLP,\,\Omega)$ are implicative.
\end{prop}

\begin{proof}
Proof is by induction.
Let a formal logic program \FLP, a literal marker $\Omega$ for \FLP, and a foundational
chain $(E_\alpha)_{\alpha\in\ORD}$ for $(\FLP,\,\Omega)$ be given.
Let an ordinal $\alpha$ be given and suppose that for all $\beta<\alpha$,
$\bigcup_{\gamma<\beta}E_\gamma\subseteq\LB\FLP\PLUS\bigcup_{\gamma<\beta}E_\gamma\RB$. There is nothing to verify if $\alpha=0$.
If $\alpha$ is a limit ordinal then it follows immediately from Lemma~\ref{lem_60}
that $\bigcup_{\beta<\alpha}E_\beta$ is included in $\LB\FLP\PLUS\bigcup_{\beta<\alpha}E_\beta\RB$.
Suppose that $\alpha$ is of the form $\delta+1$. By inductive hypothesis, $\bigcup_{\gamma<\delta}E_\gamma$ is included in $\LB\FLP\PLUS\bigcup_{\gamma<\delta}E_\gamma\RB$.
Moreover, it follows from Properties~\ref{property_50} and~\ref{property_55} that
$\BLB\FLP\PLUS\LB\FLP\PLUS\bigcup_{\gamma<\delta}E_\gamma\RB\cup E_\delta\BRB$ contains $E_\delta$.
Lemma~\ref{lem_61} then implies that
$E_\delta\subseteq\LB\FLP\PLUS\bigcup_{\gamma\leq\delta}E_\gamma\RB$.
We conclude with Property~\ref{property_56}.
\end{proof}

The next proposition will allow us to relate our framework to the well-founded semantics of a formal logic program \FLP\ either in terms of a particular foundational extensor $E$ for $(\FLP,\,\SPE{\cdot,-})$, or in terms of $\LB\FLP\PLUSDOTNEG E\RB$ for a particular foundational extensor $E$ for $(\FLP,\,\SPE{,\cdot,-})$.

\begin{prop}\label{prop_74}
Let a formal logic program \FLP, a literal marker $\Omega$ for \FLP, and a foundational extensor $E$ for $(\FLP,\,\Omega)$ be given. Then $\LB\FLP\PLUS E\RB$ is a foundational extensor for $(\FLP,\,\Omega)$.
\end{prop}

\begin{proof}
By Property~\ref{property_56}, choose an ordinal $\alpha$ and an $\alpha$-foundational chain $(E_\beta)_{\beta<\alpha}$ for $(\FLP,\,\Omega)$ with $\bigcup_{\beta<\alpha}E_\beta=E$.
Set $E_\alpha=\LB\FLP\PLUS E\RB$. Using Property~\ref{property_50}, Lemma~\ref{lem_62} and Proposition~\ref{prop_73}, it is easy to verify that $(E_\beta)_{\beta\leq\alpha}$ is an $(\alpha+1)$-foundational chain for $(\FLP,\,\Omega)$. We conclude with Property~\ref{property_56} again.
\end{proof}

As mentioned in the discussion following Example~\ref{ex_71}, our framework and the well-founded semantics of a formal logic program \FLP\ can be related using either \SPE{\cdot,-} or \SPE{-,-}; this will be a consequence of the property that follows.

\begin{property}\label{property_75}
Let a formal logic program \FLP\ and a set $E$ of negated atoms over \VOCA\ be given. Then $E$ is a foundational extensor for $(\FLP,\,\SPE{\cdot,-})$ iff $E$ is a foundational extensor for $(\FLP,\,\SPE{-,-})$.
\end{property}

We now state a counterpart to Corollary~\ref{cor_65} for foundational chains.

\begin{prop}\label{prop_76}
Let a formal logic program \FLP\ be locally consistent.
Let a literal marker $\Omega$ for \FLP\ and a set $I$ be given.
Let a set of foundational chains for $(\FLP,\,\Omega)$ of the form $\{(E_\alpha^\sigma)_{\alpha\in\ORD}\mid \sigma\in I\}$ be given.
Then $(\bigcup_{\sigma\in I}E_\alpha^\sigma)_{\alpha\in\ORD}$ is a foundational
chain for $(\FLP,\,\Omega)$
iff $\bigcup_{\sigma\in I}\bigcup_{\alpha\in\ORD}E_\alpha^\sigma$ is consistent.
\end{prop}

\begin{proof}
Only one direction of the proposition requires a proof. The argument is by induction.
For all ordinals $\alpha$, set $F_\alpha=\bigcup_{\sigma\in I}E_\alpha^\sigma$.
Assume that $\bigcup_{\alpha\in\ORD}F_\alpha$ is consistent.
Let an ordinal $\alpha$ be given, and assume that for all $\beta<\alpha$,
$(F_\gamma)_{\gamma<\beta}$ is a $\beta$-foundational chain for $(\FLP,\,\Omega)$.
Trivially, if $\alpha=0$ or if $\alpha$ is a limit ordinal then $(F_\beta)_{\beta<\alpha}$ is an $\alpha$-foundational
chain for $(\FLP,\,\Omega)$. Assume that $\alpha$ is of the form $\delta+1$.
To complete the proof of the proposition, it is clearly sufficient to show that $F_\delta\cup\BLB(\FLP\PLUS\bigcup_{\gamma<\delta}F_\gamma)+_{\Omega+\bigcup_{\gamma<\delta}F_\gamma}F_\delta\BRB$ is consistent.
By Property~\ref{property_50}, it suffices to verify that $F_\delta\cup\LB\FLP\PLUS\bigcup_{\gamma\leq\delta}F_\gamma\RB$ is consistent. But this is an immediate consequence of the fact that by Propositions~\ref{prop_64} and~\ref{prop_73}, $\bigcup_{\gamma\leq\delta}F_\gamma$, equal to $\bigcup_{\sigma\in I}\bigcup_{\gamma\leq\delta}E^\sigma_\gamma$, is an extensor for $(\FLP,\,\Omega)$.
\end{proof}

As an application of Proposition~\ref{prop_76}, we can follow the main path in the field of logic programming, be biased towards negative information, and get the following proposition.

\begin{prop}\label{prop_77}
Let a locally consistent formal logic program \FLP\ be given.
\begin{itemize}
\item
There exists a unique $\subseteq$-maximal set $E$ of negated atoms over \VOCA\ that is a foundational extensor for $(\FLP,\,\SPE{\cdot,-})$, or equivalently, for $(\FLP,\,\SPE{-,-})$.
\item
There exists a unique $\subseteq$-maximal set $F$ of literals over \VOCA\ that is a foundational extensor for $(\FLP,\,\SPE{\cdot,-})$, or equivalently, for $(\FLP,\,\SPE{-,-})$; moreover, $F$ is equal to both $\LB\FLP\PLUSDOTNEG E\RB$ and $\LB\FLP\PLUSNEG E\RB$.
\end{itemize}
\end{prop}

\begin{proof}
The existence of a unique $\subseteq$-maximal set $E$ of negated atoms over \VOCA\ that is a foundational extensor for $(\FLP,\,\SPE{\cdot,-})$, or equivalently, for $(\FLP,\,\SPE{-,-})$, follows immediately from Proposition~\ref{prop_76} and Property~\ref{property_75}. Let $\Omega$ denote either \SPE{\cdot,-} or \SPE{-,-}.
By Proposition~\ref{prop_74}, $\LB\FLP\PLUS E\RB$ is a foundational extensor for $(\FLP,\,\Omega)$. 
Let $(F_\alpha)_{\alpha\in\ORD}$ be a foundational chain for $(\FLP,\,\Omega)$.
For all ordinals $\alpha$, let $G_\alpha$ be the set of negated atoms in $F_\alpha$. We show that $(G_\alpha)_{\alpha\in\ORD}$ is a foundational chain for $(\FLP,\,\Omega)$.
Proof is by induction, so let $\alpha\in\ORD$ be given, and assume that for all $\beta<\alpha$, $(G_\gamma)_{\gamma<\beta}$ is a $\beta$-foundational chain for $(\FLP,\,\Omega)$. Trivially, if $\alpha=0$ or $\alpha$ is a limit ordinal then $(G_\beta)_{\beta<\alpha}$ is an $\alpha$-foundational chain for $(\FLP,\,\Omega)$. Suppose that $\alpha$ is of the form $\delta+1$. Obviously, $\LB\FLP\PLUS\bigcup_{\beta<\delta}F_\beta\RB=\LB\FLP\PLUS\bigcup_{\beta<\delta}G_\beta\RB$. This together with the fact that $F_\delta$ is a supporting extensor for
$\bigl(\LB\FLP\PLUS\bigcup_{\beta<\delta}F_\beta\RB,\,\Omega+\bigcup_{\beta<\delta}F_\beta\bigr)$ implies immediately that $G_\delta$ is a supporting extensor for $\bigl(\LB\FLP\PLUS\bigcup_{\beta<\delta}G_\beta\RB,\,\Omega+\bigcup_{\beta<\delta}G_\beta\bigr)$, which completes the proof that $(G_\alpha)_{\alpha\in\ORD}$ is a foundational chain for $(\FLP,\,\Omega)$. Obviously, $\LB\FLP\PLUS\bigcup_{\alpha\in\ORD}G_\alpha\RB=\LB\FLP\PLUS\bigcup_{\alpha\in\ORD}F_\alpha\RB$. Moreover, $\bigcup_{\alpha\in\ORD}G_\alpha$ is a subset of $E$. Hence $F$, which is a subset of $\LB\FLP\PLUS\bigcup_{\alpha\in\ORD}F_\alpha\RB$ by Proposition~\ref{prop_73}, is included in $\LB\FLP\PLUS E\RB$ by Lemma~\ref{lem_60}.
Hence $\LB\FLP\PLUS E\RB$ is the unique $\subseteq$-maximal set $F$ of literals over \VOCA\ that is a foundational extensor for $(\FLP,\,\Omega)$.
\end{proof}

\subsection{Relationship to the well-founded semantics}

The well-founded semantics takes the class of sets of positive rules as object of study; so again, it is legitimate to study the well-founded semantics on the basis of the class of
symmetric formal logic programs. But we will see that the hypothesis of symmetry is unnecessarily strong: it is enough to focus on locally consistent formal logic programs. If one assumes that \VOCA\ is equal to \VOCAB, and if one remains in the realm of symmetric formal logic programs, then Definition~\ref{defn_78} captures the notion of well-founded model. Here not assuming that \VOCA\ and \VOCAB\ are equal offers a genuine generalisation.

\begin{defn}\label{defn_78}
Let a formal logic program \FLP\ be given. Define two sequences $(E_\alpha^+)_{\alpha\in\ORD}$ and $(E_\alpha^-)_{\alpha\in\ORD}$ of sets of literals as follows. Let an ordinal $\alpha$ be given, and assume that $E_\beta^+$ and $E_\beta^-$ have been defined for all $\beta<\alpha$.
\begin{itemize}
\item
$E_\alpha^+$ is defined as the set of closed instances of the $\subseteq$-smallest set $X$ of atoms over \VOCA\ such that for all $\psi\in X$, $\bigcup_{\beta<\alpha}E_\beta^-\cup X$ forces $\FLP[\psi]$.
\item
$E_\alpha^-$ is defined as the set of closed instances of the $\subseteq$-largest set $X$ of negated atoms over \VOCA\ such that for all $\psi\in X$, $\bigcup_{\beta<\alpha}E_\beta^+\cup X$ forces $\FLP[\psi]$.
\end{itemize}
Set $E=\bigcup_{\alpha\in\ORD}(E_\alpha^+\cup E_\alpha^-)$.
If $E$ is a partial interpretation (is consistent), then \FLP\ is said to \emph{have a well-founded model} and $E$ is called the \emph{well-founded model of\/ \FLP}.
\end{defn}

\begin{property}\label{property_79}
Let a formal logic program $\FLP=(\varphi_\wp^\epsilon)_{(\wp,\epsilon)\in\IND}$ be given. Let $(E_\alpha^+)_{\alpha\in\ORD}$ and $(E_\alpha^-)_{\alpha\in\ORD}$ be the two sequences of sets of literals defined in Definition~\ref{defn_78}. Then for all ordinals $\alpha$,
$\bigl\{\selects{\bigcup_{\beta<\alpha}E_\beta^-}{-}{\varphi_\wp^+}\rightarrow\wp(v_1,\dotsc,v_n)\bigm| n\in\NSET,\,\wp\in\PRDVN{n}\bigr\}\WMODELS E^+_\alpha$.
\end{property}

Recall that by Proposition~\ref{prop_77}, we can talk about `the $\subseteq$-maximal foundational extensor for $(\FLP,\,\SPE{\cdot,-})$' when \FLP\ is locally consistent.
The next proposition shows that this extensor fully characterises the notion of well-founded model. The proposition does more than embed the well-founded semantics into our framework as it encompasses all formal logic programs that are locally consistent rather than just symmetric, and as it does not assume that \VOCA\ and \VOCAB\ are equal.

\begin{prop}\label{prop_80}
Let \FLP\ be a locally consistent formal logic program, and let $F$ be the $\subseteq$-maximal foundational extensor for $(\FLP,\,\SPE{\cdot,-})$.
Then \FLP\ has a well-founded model, which is precisely the set of closed instances of members of $F$.
\end{prop}

\begin{proof}
Let $(E_\alpha^+)_{\alpha\in\ORD}$ and $(E_\alpha^-)_{\alpha\in\ORD}$ be the sequences of literals defined in Definition~\ref{defn_78}. For all ordinals $\alpha$, let $D_\alpha^+$ be the set of atoms over \VOCA\ all of whose closed instances belong to $E_\alpha^+$, and let $D_\alpha^-$ be the set of negated atoms over \VOCA\ all of whose closed instances belong to $E_\alpha^-$.
Note that for all $\alpha\in\ORD$, $E_\alpha^+$ and $E_\alpha^-$ are the sets of closed instances of members of $D_\alpha^+$ and $D_\alpha^-$, respectively.
Set $E=\bigcup_{\alpha\in\ORD}(E_\alpha^+\cup E_\alpha^-)$.
By Proposition~\ref{prop_77}, let $(F_\alpha)_{\alpha\in\ORD}$ be a foundational chain for $(\FLP,\,\SPE{\cdot,-})$ such that all members of $\bigcup_{\alpha\in\ORD}F_\alpha$ are negated atoms and $F=\LB\FLP\PLUSDOTNEG\bigcup_{\alpha\in\ORD}F_\alpha\RB$.

We first show that for all ordinals $\alpha$, $(D_\beta^-)_{\beta<\alpha}$ is an $\alpha$-foundational chain for $(\FLP,\,\SPE{\cdot,-})$. Proof is by induction, so let ordinal $\alpha$ be given, and assume that for all $\beta<\alpha$, $(D_\gamma^-)_{\gamma<\beta}$ is a $\beta$-foundational chain for $(\FLP,\,\SPE{\cdot,-})$. Trivially, if $\alpha=0$ or $\alpha$ is a limit ordinal then $(D_\beta^-)_{\beta<\alpha}$ is an $\alpha$-foundational chain for $(\FLP,\,\SPE{\cdot,-})$. Suppose that $\alpha$ is of the form $\delta+1$. Let $\psi\in D_\delta^-$ be given. Then $\bigcup_{\beta<\delta}E_\beta^+\cup D_\delta^-$ forces $\FLP[\psi]$. Together with Property~\ref{property_79}, this implies that $\LB\FLP\PLUSDOTNEG\bigcup_{\beta<\delta}D_\beta^-\RB\cup D_\delta^-$ forces $\FLP[\psi]$, hence also $(\FLP\PLUSDOTNEG\bigcup_{\beta<\delta}D_\beta^-)[\psi]$, which together with Corollary~\ref{cor_53} implies that $(D_\beta^-)_{\beta<\alpha}$ is an $\alpha$-foundational chain for $(\FLP,\,\SPE{\cdot,-})$, as wanted. Now by the definition of $F$ and Proposition~\ref{prop_74},
$\LB\FLP\PLUSDOTNEG\bigcup_{\alpha\in\ORD}D_\alpha^-\RB$ is included in $F$. We conclude with Lemma~\ref{lem_60}, Property~\ref{property_79} again and Property~\ref{property_45} that the set of closed instances of members of $F$ contains $E$, which is therefore consistent. Hence \FLP\ has a well-founded model, which is $E$.

To establish the converse, we show by induction that for all ordinals $\alpha$,
\begin{itemize}
\item
all closed instances of members of $F_\alpha$ belong to $E$, and
\item
all closed instances of members of $\LB\FLP\PLUSDOTNEG\bigcup_{\beta<\alpha}F_\beta\RB$ belong to $E$.
\end{itemize}
So let an ordinal $\alpha$ be given and assume that (i) for all ordinals $\beta<\alpha$, all closed instances of members of $F_\beta$ belong to $E$, and (ii) for all ordinals $\beta<\alpha$, all closed instances of members of $\LB\FLP\PLUSDOTNEG\bigcup_{\gamma<\beta}F_\gamma\RB$ belong to $E$.
Note the following:
\begin{center}
$(\star)$\quad for all literals $\psi$ over \VOCA, $(\FLP\PLUSDOTNEG\bigcup_{\beta<\alpha}F_\beta)[\psi]\cup\bigcup_{\beta<\alpha}F_\beta$ forces $\FLP[\psi]$.
\end{center}
Let an ordinal $\delta$ be such that $E=E^+_\delta\cup E^-_\delta$. Using $(\star)$, we obtain by induction that for all $\gamma\in\ORD$ and $\psi\in\LB\FLP\PLUSDOTNEG\bigcup_{\beta<\alpha}F_\beta\RB_\gamma$, if $\bigcup_{\beta'<\gamma}\LB\FLP\PLUSDOTNEG\bigcup_{\beta<\alpha}F_\beta\RB_{\beta'}$ forces $(\FLP\PLUSDOTNEG\bigcup_{\beta<\alpha}F_\beta)[\psi]$ then $\bigcup_{\beta'<\gamma}\LB\FLP\PLUSDOTNEG\bigcup_{\beta<\alpha}F_\beta\RB_{\beta'}\cup\bigcup_{\beta<\alpha}F_\beta$ forces $\FLP[\psi]$; this together with (i) easily implies that if $X^+$ and $X^-$ denote the set of atoms and the set of negated atoms in $\bigcup_{\beta'<\gamma}\LB\FLP\PLUSDOTNEG\bigcup_{\beta<\alpha}F_\beta\RB_{\beta'}\cup\{\psi\}$, respectively, then
\begin{itemize}
\item
the set of closed instances of members of $X^+$ is included in $E^+_{\delta+1}$, and
\item
the set of closed instances of members of $X^-\cup\bigcup_{\beta<\alpha}F_\beta$ is included in $E^-_{\delta+1}$.
\end{itemize}
Hence all closed instances of members of $\LB\FLP\PLUSDOTNEG\bigcup_{\beta<\alpha}F_\beta\RB$ belong to $E$.
Using $(\star)$ again, we obtain that for all $\psi\in F_\alpha$,
since $\LB\FLP\PLUSDOTNEG\bigcup_{\beta<\alpha}F_\beta\RB\cup F_\alpha$ forces $(\FLP\PLUSDOTNEG\bigcup_{\beta<\alpha}F_\beta)[\psi]$ by Corollary~\ref{cor_53}, then $E\cup F_\alpha$ forces $\FLP[\psi]$; this easily implies that the set consisting of the closed instances of either the negated atoms in $E$ or the members of $F_\alpha$ is included in $E^-_{\delta+1}$. Hence all closed instances of members of $F_\alpha$ belong to $E$. Since $F$ is equal to $\LB\FLP\PLUSDOTNEG\bigcup_{\alpha\in\ORD}F_\alpha\RB$, we have shown that all closed instances of members of $F$ belong to $E$, which completes the proof of the proposition.
\end{proof}

\section{Conclusion}

Given a formal logic program \FLP, we have defined the set $\LB\FLP\RB$ of literals generated by \FLP\ following a process that can be intuitively described as: fire the rules in \FLP\ transfinitely often, and at each stage interpret disjunction and existential quantification constructively to determine whether an instance of the body of a rule should be activated, the rule fired, and the corresponding instance of the head added to $\LB\FLP\RB$. The view that has been adopted is that $\LB\FLP\RB$ captures the operational semantics of \FLP. This view is closely related to Kripke-Kleene semantics (this is the contents of Proposition~\ref{prop_36}). We have introduced the notion of `literal marker for \FLP' to formalise the intuitive idea of `marking some literals in the bodies of some rules in \FLP'. Given such a literal marker $\Omega$ for \FLP\ and a set $E$ of literals conceived of as a collection of hypotheses, meant to be assumed only in the contexts authorised by $\Omega$, we have formalised the intuitive operation of making these contextual, local assumptions, resulting in a new formal logic program, denoted $\FLP\PLUS E$; the denotational semantics of that program is of course captured by $\LB\FLP\PLUS E\RB$. For a given literal marker $\Omega$ for \FLP\ and a given set $E$ of literals, $\LB\FLP\PLUS E\RB$ can also be conceived of as an alternative semantics to \FLP, and we have seen how to choose $\Omega$ and $E$ in order to retrieve the answer-set, the stable model and the well-founded semantics.
\begin{itemize}
\item
Answer-sets are captured by sets of the form $\LB\FLP\PLUS E\RB$ in which $\Omega$ marks the occurrences of literals of the form $\neg\mathit{atom}$, represented in the usual setting as $\NOT\mathit{atom}$, or of the form $\mathit{atom}$, represented in the usual setting as $\NOT\neg\mathit{atom}$, and $E$ is a maximal (in a strong sense) set of literals that $\FLP\PLUS E$ does not refute (this is the contents of Proposition~\ref{prop_68}).
\item
Stable models are captured by sets of the form $\LB\FLP\PLUS E\RB$ in which $\Omega$ marks all occurrences of negated atoms in the bodies of all rules, and $E$ is a maximal set of negated atoms which determines a complete set of literals that $\FLP\PLUS E$ confirms (this is the contents of Proposition~\ref{prop_70}).
\item
The well-founded model is the set $\LB\FLP\PLUS E\RB$ in which $\Omega$ marks all occurrences of negated atoms in the bodies of all negative rules, and $E$ is the maximal set of negated atoms that $\FLP\PLUS E$ confirms in a strong sense, based on the concept of a set of hypotheses that can get `self-confirmation' with no additional help but what can be derived from \FLP\ itself, put into action transfinitely often (this is the contents of Propositions~\ref{prop_77} and~\ref{prop_80}).
\end{itemize}
The relationships have actually been established for a class of logic programs more general than those usually considered in the literature, but for which those semantics could be naturally adapted. The classes of extensors (legitimate sets of hypotheses) that have been introduced can be subjected to natural variations; the choices for $\Omega$ can range from fully biased towards negated atoms to fully biased towards nonnegated atoms, or seek some balance between both kinds of literals, etc. Hence the three semantics captured by $\LB\FLP\PLUS E\RB$ for the specific choices of $\Omega$ and $E$ that have been described are members of families of semantics determined by a pair $(\Omega,E)$ that naturally satisfies more general properties. We have not investigated these alternative semantics for lack of space, but we think that one of the main contributions of this paper is to have laid the foundation for such a work, with applications to hypothetical reasoning in knowledge-based systems, where hypotheses are applied locally and contextually, and are constrained to satisfy variations on properties such a confirmation or nonrefutation.

Though Kripke-Kleene, the answer-set, the stable model and the well-founded semantics are expressed in terms of `intended' or `preferred' models, we do not view $\LB\FLP\RB$ as the intended model of what we have called the classical logical form, denoted $\CLF(\FLP)$, of the formal logic program \FLP.
Indeed, we have carefully not defined a formal logic program as a set of logical formulas. We have chosen to model the behavior of a set of rules that can fire transfinitely often, hence provide an operational semantics, which does not require to represent rules as logical implications. Another paper will present a declarative semantics, in such a way that $\bigl\{\Box\varphi\mid\varphi\in\LB\FLP\RB\bigr\}$ is precisely the set of formulas of the form $\Box\varphi$ with $\varphi$ a literal that are logical consequences of a set of modal formulas that is obtained from $\CLF(\FLP)$ by preceding all occurrences of literals with the modal operator of necessity. (The main work is to capture properly the transformation of \FLP\ into $\FLP\PLUS E$---marking literals has to find its logical counterpart---, and to properly represent the hypotheses). In this setting, `logical consequence' is interpreted classically, that is, in reference to a notion of interpretation that generalises the interpretations used in epistemic logic, in which every formula is either true or false (not undefined), negation is classical, and the law of excluded middle holds but is irrelevant, because a rule such as $q\leftarrow p\vee\neg p$ is logically translated into $\Box p\vee\Box\neg p\rightarrow\Box q$: to derive $q$, derive $p$ or derive $\neg p$, and $q\leftarrow p\vee\neg p$ does not automatically fire because $\Box p\vee\Box\neg p$ is not valid.

\bibliographystyle{acmtrans}

\end{document}